\documentclass[11pt]{article}

\usepackage{latexsym,amsmath}

\newcounter{subequation}[equation]
\makeatletter

\def\bcite{\@ifnextchar [{\@tempswatrue\@bcitex}{\@tempswafalse\@bcitex[]}}
\def\@bcitex[#1]#2{\if@filesw\immediate\write\@auxout{\string\citation{#2}}\fi
  \let\@bcitea\@empty
  \@bcite{\@for\@bciteb:=#2\do
    {\@bcitea\def\@bcitea{,\penalty\@m\ }%
     \def\@tempa##1##2\@nil{\edef\@bciteb{\if##1\space##2\else##1##2\fi}}%
     \expandafter\@tempa\@bciteb\@nil
     \@ifundefined{b@\@bciteb}{{\reset@font\bf ?}\@warning
       {Citation `\@bciteb' on page \thepage \space undefined}}%
     \hbox{\csname b@\@bciteb\endcsname}}}{#1}}
\def\@bcite#1#2{{#1\if@tempswa , #2\fi}}

\def\thesubequation{\theequation\@alph\c@subequation}
\def\@subeqnnum{{\rm (\thesubequation)}}
\def\slabel#1{\@bsphack\if@filesw {\let\thepage\relax
   \xdef\@gtempa{\write\@auxout{\string
      \newlabel{#1}{{\thesubequation}{\thepage}}}}}\@gtempa
   \if@nobreak \ifvmode\nobreak\fi\fi\fi\@esphack}
\def\subeqnarray{\stepcounter{equation}
\let\@currentlabel=\theequation\global\c@subequation\@ne
\global\@eqnswtrue
\global\@eqcnt\z@\tabskip\@centering\let\\=\@subeqncr
$$\halign to \displaywidth\bgroup\@eqnsel\hskip\@centering
  $\displaystyle\tabskip\z@{##}$&\global\@eqcnt\@ne
  \hskip 2\arraycolsep \hfil${##}$\hfil
  &\global\@eqcnt\tw@ \hskip 2\arraycolsep
  $\displaystyle\tabskip\z@{##}$\hfil
   \tabskip\@centering&\llap{##}\tabskip\z@\cr}
\def\endsubeqnarray{\@@subeqncr\egroup
                     $$\global\@ignoretrue}
\def\@subeqncr{{\ifnum0=`}\fi\@ifstar{\global\@eqpen\@M
    \@ysubeqncr}{\global\@eqpen\interdisplaylinepenalty \@ysubeqncr}}
\def\@ysubeqncr{\@ifnextchar [{\@xsubeqncr}{\@xsubeqncr[\z@]}}
\def\@xsubeqncr[#1]{\ifnum0=`{\fi}\@@subeqncr
   \noalign{\penalty\@eqpen\vskip\jot\vskip #1\relax}}
\def\@@subeqncr{\let\@tempa\relax
    \ifcase\@eqcnt \def\@tempa{& & &}\or \def\@tempa{& &}
      \else \def\@tempa{&}\fi
     \@tempa \if@eqnsw\@subeqnnum\refstepcounter{subequation}\fi
     \global\@eqnswtrue\global\@eqcnt\z@\cr}
\let\@ssubeqncr=\@subeqncr
\@namedef{subeqnarray*}{\def\@subeqncr{\nonumber\@ssubeqncr}\subeqnarray}
\@namedef{endsubeqnarray*}{\global\advance\c@equation\m@ne%
                           \nonumber\endsubeqnarray}

\makeatother

\DeclareFontFamily{OT1}{rsfs10}{}
\DeclareFontShape{OT1}{rsfs10}{m}{n}{ <-> rsfs10 }{}
\DeclareMathAlphabet{\mathscript}{OT1}{rsfs10}{m}{n}

\numberwithin{equation}{section}

\newcommand{\ns}{\normalsize}

\newcommand{\be}{\begin{equation}}
\newcommand{\ee}{\end{equation}}
\newcommand{\nn}{\nonumber}
\newcommand{\bea}{\begin{eqnarray}}
\newcommand{\eea}{\end{eqnarray}}
\newcommand{\bsea}{\begin{subeqnarray}} 
\newcommand{\esea}{\end{subeqnarray}}

\newcommand{\tr}{\textrm{tr}}
\newcommand{\Tr}{\textrm{Tr}}

\def\eg{{\it e.g.}\ }

\def\a{\alpha}
\def\b{\beta}
\def\g{\gamma}
\def\c{\chi}
\def\d{\delta}
\def\e{\epsilon}

\def\f{\phi}

\def\k{\kappa}
\def\l{\lambda}
\def\m{\mu}
\def\n{\nu}
\def\o{\omega}
\def\p{\pi}

\def\r{\rho}
\def\s{\sigma}

\def\x{\xi}

\def\w{\wedge}

\def\L{\Lambda}
\def\O{\Omega}

\def\cA{{\cal A}}
\def\cF{{\cal F}}

\def\cC{{\cal C}}

\def\w{\wedge}
\def\cG{{\cal G}}
\def\cH{{\cal H}}
\def\bcG{{\bar{\cal G}}}
\def\bcH{{\bar{\cal H}}}
\def\bcF{{\bar{\cal F}}}
\def\tcF{{\tilde{\cal F}}}
\def\tcA{{\tilde{\cal A}}}
\def\cB{{\cal B}}
\def\cJ{{\cal J}}
\def\cN{{\cal N}}
\def\cD{{\cal D}}

\begin{document}

%%%%%%%%%%%%%%%%%%%%%%%%%%%%%%%%%%%%%%%%%%%%%%%%%%%%%%%%%%%%%%%%%%%%%%

\begin{titlepage}

\vspace{-3cm}

\title{
   \hfill{\ns OUTP-99-49P, Imperial/TP/99--0/12\\}
   \hfill{\ns hep-th/9911156\\[.5cm]}
   {\LARGE Heterotic Anomaly Cancellation in Five Dimensions}}
\author{
   Andr\'e Lukas$^1$ and K.S. Stelle$^2$\\[0.5cm]
   {\ns $^1$Department of Physics, Theoretical Physics, 
       University of Oxford} \\
   {\ns 1 Keble Road, Oxford OX1 3NP, United Kingdom}\\[0.3cm]
   {\ns $^2$The Blackett Laboratory, Imperial College, London SW7 2BZ, UK}} 
\date{}

\maketitle

\begin{abstract}
We study the constraints on five--dimensional $\cN =1$ heterotic
M--theory imposed by a consistent anomaly--free coupling of bulk and boundary
theory. This requires analyzing the cancellation of
triangle gauge anomalies on the four--dimensional orbifold planes due
to anomaly inflow from the bulk. We find that the semi--simple part of
the orbifold gauge groups and certain $U(1)$ symmetries have to be free of
quantum anomalies. In addition there can be several anomalous $U(1)$
symmetries on each orbifold plane whose anomalies are cancelled by a
non--trivial variation of the bulk vector fields. The mixed
$U(1)$ non--abelian anomaly is universal and there is at most one
$U(1)$ symmetry with such an anomaly on each plane. 
In an alternative approach, we also analyze the coupling of
five--dimensional gauged supergravity to orbifold gauge theories.
We find a somewhat generalized structure of anomaly cancellation in
this case which allows, for example, non--universal mixed $U(1)$ gauge
anomalies. Anomaly cancellation from the perspective
of four--dimensional $\cN =1$ effective actions obtained
from $E_8\times E_8$ heterotic string-- or M--theory by reduction
on a Calabi--Yau three--fold is studied as well. The results are
consistent with the ones found for five--dimensional heterotic
M--theory. Finally, we consider some related issues of
phenomenological interest such as model building with anomalous $U(1)$
symmetries, Fayet--Illiopoulos terms and threshold corrections to
gauge kinetic functions.
\end{abstract}

\thispagestyle{empty}

\end{titlepage}

%%%%%%%%%%%%%%%%%%%%%%%%%%%%%%%%%%%%%%%%%%%%%%%%%%%%%%%%%%%%%%%%%%%%%%%

\section{Introduction}

%%%%%%%%%%%%%%%%%%%%%%%%%%%%%%%%%%%%%%%%%%%%%%%%%%%%%%%%%%%%%%%%%%%%%%%

Green--Schwarz type anomaly cancellation of gauge and gravitational
quantum anomalies~\cite{gs} is one of the key ingredients of
string theory. Its role became even more pronounced in the wider
context of M--theory, particularly in situations where branes are
coupled to higher--dimensional bulk theories. In those cases, quantum
anomalies that may arise on the brane worldvolume are cancelled by an
anomalous variation of the bulk theory which is of Green--Schwarz type.
This mechanism is also referred to as anomaly inflow~\cite{ch,bh} from
the bulk into the brane. A prominent example is given by the M--theory
five--brane. Once coupled to 11--dimensional supergravity, its
quantum anomaly is cancelled by anomaly inflow induced by the
variation of a Green--Schwarz term that has to be added to
11--dimensional supergravity~\cite{dlm}. In an earlier
work~\cite{ddf}, a similar mechanism has been studied for the
five--brane of the $SO(32)$ heterotic string. Another example, that is
of particular relevance for the present paper is provided by the
Ho\v rava--Witten construction~\cite{hw1,hw2} of strongly coupled
$E_8\times E_8$ heterotic string. In this construction,
11--dimensional supergravity on the orbifold $S^1/Z_2$ is considered.
The quantum anomalies of the $E_8$ gauge fields residing on the
two 10--dimensional orbifold fixed planes as well as gravitational
anomalies arising on those planes are cancelled by anomaly inflow from
11--dimensional supergravity. 

Green--Schwarz anomaly cancellation mechanisms are particularly relevant from
two different perspectives. First, they provide a tool to construct
consistent anomaly--free theories, particularly those in which branes
are coupled to bulk theories. In fact, anomaly cancellation has been
the main guidance in the construction of 11--dimensional supergravity
on the orbifold $S^1/Z_2$~\cite{hw1,hw2} as well as on other
orbifolds~\cite{dm,w,s,s1,flo}. Secondly, Green--Schwarz anomaly
cancellation or certain remnants thereof arising in low--energy
effective theories associated with string-- or M--theory
compactifications are of considerable phenomenological importance.
Correspondingly, such low--energy cancellation
mechanisms, that is the ``anomalous'' $U(1)$ symmetries that typically
arise in this context, have been investigated early on, particularly
in heterotic theories~\cite{dsww,dsw,lns,ads,dis,dg,dl}. A more
recent analysis for string orbifolds can be found in Ref.~\cite{kn}. 

Aspects of anomalous $U(1)$ symmetries in the context of heterotic
M--theory have been addressed in Ref.~\cite{mr,lpt,bddr}. Particularly,
Ref.~\cite{lpt} contains an interesting discussion of some of the
issues that become apparent in the strongly coupled limit. It is known
that models based on type I string theory may have a rich structure of
low--energy anomaly cancellation~\cite{sag}. Recently, there is
considerable interest in this phenomenon~\cite{iru,p,lln,iru1,iq} and
its relation to heterotic models~\cite{lln}.

\vspace{0.4cm}

The purpose of the present paper is to investigate systematically low--energy
anomaly cancellation in the context of the $E_8\times E_8$ heterotic string-- or
M--theory. That is, we will determine the structure of anomalous $U(1)$
symmetries both from the viewpoint of the four-- and five--dimensional effective
actions. This will be done in a three--fold way. In a first step, we will
investigate anomaly cancellation in five--dimensional heterotic
M--theory~\cite{losw1,elpp,losw2}. This theory is obtained from its
11--dimensional counterpart (that is, the Ho\v rava--Witten construction) by
reduction on a Calabi--Yau three--fold with a non--zero mode of the
antisymmetric tensor field and general vector bundles. Properties of such
non--standard embedding vacua with general bundles, within the context of the
strongly coupled heterotic string, have been considered in
Ref.~\cite{benakli1,stieb,lpt,nse}. The five--dimensional effective actions
associated to such vacua have been analyzed in Ref.~\cite{nse}. These theories
consist of a five--dimensional $\cN =1$ gauged bulk supergravity coupled to two
four--dimensional orbifold planes that carry $\cN=1$ gauge theories. Within this
setting we will study the cancellation of quantum (gauge) anomalies on the
orbifold planes induced by anomaly inflow from the five--dimensional bulk
supergravity theory. In other word, we are analyzing, within the framework of
heterotic M--theory, the set of four--dimensional gauge theories that can be
consistently coupled to gauged five--dimensional $\cN =1$ supergravity. 

In a second step, we will study a similar question in a somewhat more general
context. We will consider plain 11--dimensional supergravity reduced on a
Calabi--Yau three fold in the presence of an internal non--zero mode of the
antisymmetric tensor field. The resulting five--dimensional theory is again a
gauged $\cN =1$ supergravity similar to the bulk part of five--dimensional
heterotic M--theory. Then we shall consider this five--dimensional theory on an
$S^1/Z_2$ orbifold, viewed as a background solution to the five-dimensional
supergravity equations. By the use of anomaly cancellation, we shall investigate
which four--dimensional $\cN =1$ gauge theories (arising as ``twisted states'' on
the orbifold fixed planes) can be consistently coupled to this five--dimensional
supergravity theory. Note that in this approach, in contrast to the previous
one, the orbifold planes and their particle content are not necessarily
inherited from the 11 dimensional Ho\v rava--Witten construction. It is
therefore more general and can be expected to lead to a wider class of models. 

Both the above approaches are aimed at finding consistent five--dimensional
M--theory models coupled to four--dimensional gauge theories by using the
constraints arising from anomaly cancellation. Clearly, these models are of
considerable phenomenological interest as they may lead to potentially realistic
$\cN =1$ supergravity models in four dimensions. Finally, to complement the
five--dimensional approaches and to  investigate more closely the
phenomenological aspects, we will perform a similar analysis directly for the
four--dimensional effective actions associated with the
$E_8\times E_8$ heterotic string-- or M--theory reduced on $S^1$ times
Calabi--Yau three--folds. That is, we will analyze the structure of anomalous
$U(1)$ symmetries from a four--dimensional viewpoint.

Given the considerable literature already existing on anomalous $U(1)$ symmetries
in heterotic theories, what is the motivation for our present study? First of
all, we are interested in the new five--dimensional aspects of the problem such
as anomaly inflow. Moreover, the five--dimensional picture appears to
immediately run into conflict with certain assumptions that are usually made
about anomalous $U(1)$ symmetries in the $E_8\times E_8$ heterotic theory. Let
us cite some of these standard assumptions. The starting point is the assertion
that the various $U(1)$ quantum anomalies are cancelled by a shift of the dilaton
$S$. Due to its universal coupling, there is at most one anomalous
$U(1)$ symmetry which has the same mixed gauge anomaly coefficients for all
gauge factor (and, in fact, for gravity, as well) in the observable and the
hidden sector. This implies that there have to be matter fields charged under
this $U(1)$ symmetry in both sectors. In this sense, the anomalous $U(1)$
symmetry ``extends'' over the observable and the hidden sector. How can such a
$U(1)$ symmetry arise in a five--dimensional theory in which the hidden and the
observable sector are confined to four--dimensional brane worldvolumes that are
spatially separated by a ``gravity only'' bulk? In this paper, we attempt to resolve
this puzzle as well as related ones, thereby establishing a consistent picture of
$E_8\times E_8$ low--energy anomaly cancellation.

\vspace{0.4cm}

In the next section we begin by deriving the effective five--dimensional action
of heterotic M--theory. While some of the results have already been obtained in
the literature~\cite{losw1,elpp,losw2,nse}, we will focus on those aspects that
are particularly relevant for our purposes. Specifically, we will derive all
parts of the effective action that are related to bulk antisymmetric tensor
fields. Care will be taken, in this derivation, to incorporate the most general
structure of gauge fields on the orbifold planes. As a result, we obtain the
most general form of bulk Chern--Simons terms as well as Bianchi identities. The
latter controls the coupling between the bulk antisymmetric tensor fields and the
gauge fields on the orbifold planes. We will find it useful to split the
associated low--energy gauge groups $\cG_n$, where $n=1,2$ numbers the orbifold
planes, into two parts, that is, $\cG_n=\cH_n\times\cJ_n$.  The first part,
$\cH_n$ contains the semi--simple part of the gauge group and certain $U(1)$
factors. We will describe these $U(1)$ factors as being of type II. They are
characterized by the property that the $U(1)$ factor is not part of the internal
bundle structure group. On the other hand, $\cJ_n$ contains $U(1)$ factors that
we shall describe as being of type I. Their defining property is that the $U(1)$
symmetry is part of the internal bundle structure group. This distinction
between two types of $U(1)$ symmetries is not new~\cite{dsww,dg}. However, it is
frequently not taken into account explicitly. 

In section 3, we shall study the classical variation of the five--dimensional
action under orbifold gauge symmetries. This classical variation arises due to a
non--trivial gauge transformation law of the bulk vector fields residing in the
$\cN =1$ gravity and vector multiplets. An important role is also played by a
certain bulk Chern--Simons term that originates from the internal non--zero mode
of the anti--symmetric tensor field and is closely related to the gauging of the
five--dimensional supergravity theory~\cite{losw1,losw2}. Carrying out an index
theorem calculation~\cite{w1,gsw2}, we check explicitly that this classical
variation cancels the quantum variation on the orbifold planes due to triangle
diagrams. As a result, we obtain the triangle anomaly coefficients in terms of
topological data of the underlying Calabi--Yau compactification.  This implies
the following general structure of anomalies. First of all, anomaly cancellation
works independently on each orbifold plane due to anomaly inflow from the bulk.
Further, the
$\cH_n$ parts of the gauge groups are quantum anomaly--free, that is, cubic
anomalies within $\cH_n$ and mixed gravitational anomalies of $\cH_n$ gauge
fields vanish. For the semi-simple part of $\cH_n$ this result is
well--known~\cite{w1}. In addition, we conclude that the type II $U(1)$
symmetries in $\cH_n$ are anomaly--free in above sense. The remaining $U(1)$
symmetries of type I in $\cJ_n$ may have quantum anomalies that are cancelled by
anomaly inflow. The associated anomaly coefficients for the hidden and the
observable sector are generically unrelated. On each orbifold plane, the anomaly
coefficients for the mixed $U(1)$ gauge anomalies (diagrams with one type I
$U(1)$ gauge field and two gauge fields from $\cH_n$), the mixed $U(1)$
gravitational anomalies (diagrams with one type I $U(1)$ gauge field and two
gravitons) and the cubic $U(1)$ anomalies (diagrams with three type I $U(1)$
gauge fields) can be non--zero and are likewise generically unrelated. However,
the mixed $U(1)$ gauge anomaly is the same for each factor within $\cH_n$ and is
in this sense universal. This implies that, in each sector, there is at most one
type I $U(1)$ symmetry with all three types of anomalies non--vanishing. In
addition, there can be other type I
$U(1)$ symmetries with vanishing mixed gauge anomaly but non--vanishing
gravitational and cubic anomaly. We stress again that those statements apply
independently two both sectors. There could, for example, be two ``anomalous''
$U(1)$ symmetries, one in each sector, both with mixed gauge anomalies but
unrelated anomaly coefficients. As another example, there could be an anomalous
$U(1)$ symmetry in the observable sector while the hidden sector is
non--anomalous.

In section 4, we will study five--dimensional anomaly cancellation in the
somewhat more general context described above. While, of course, we reproduce
all the models that were found in the context of five--dimensional heterotic
M--theory, there are also some generalizations. Most notably, the total size of
the orbifold gauge groups $\cG_n$ is not restricted to fit into $E_8$ and the
mixed $U(1)$ gauge anomaly can also depend on the factor in $\cH_n$, that is, it
can be non--universal. Although those generalizations can be realized by
coupling four--dimensional gauge theories to a special version of
five--dimensional supergravity obtained from M--theory, it is not clear whether
they are actually part of M--theory. One may attempt to realize such models in
the framework of heterotic M--theory on Calabi--Yau three--folds with
five--branes~\cite{nse,dlow,dlow1,lows}. Although this is of considerable
phenomenological interest we will not explicitly address this issue here.

In section 5, we derive the relevant parts of the four--dimensional effective
actions associated with $E_8\times E_8$ heterotic string-- or M--theory on $S^1$
times a Calabi--Yau three--fold. Again, we study the classical variation of this
action to find the pattern of anomalous $U(1)$ symmetries. As required on general
grounds, the results turn out to be consistent with the ones obtained for
five--dimensional heterotic M--theory. We also gain some more insight into the
four--dimensional mechanism that leads to Green--Schwarz cancellation. In the
standard gauge kinetic functions for $\cH_n$ given by $f_n=S\mp\e_S\b_iT^i$ it
is a shift of the $T^i$ moduli rather than the dilaton $S$ that leads to anomaly
cancellation. This resolves the puzzle raised above, since the standard
assumptions about anomalous $U(1)$ symmetries are based on a cancellation induced
by a dilaton shift. In fact, the confusion is precisely one between the type I
and type II $U(1)$ symmetries. Anomalous type II (as well as presumably type I)
$U(1)$ symmetries arise in models originating from the $SO(32)$ heterotic theory
on a Calabi-Yau three--fold. In those cases, the type II anomaly cancellation
works via a shift of the dilaton and, hence, the $U(1)$ symmetry should have the
universal properties that are commonly assumed. However, in models originating
from the $E_8\times E_8$ heterotic theory on a Calabi--Yau three--fold, we have
seen that type II $U(1)$ symmetries are alway non--anomalous. The basic reason
for this difference~\cite{dg} is that $SO(32)$ has an independent fourth order
invariant while $E_8$ has not. Hence, applying the type II anomaly cancellation
patterns to $E_8\times E_8$ models is not appropriate as type II symmetries are
always anomaly--free in such models. The type I $U(1)$ symmetries that may be
anomalous in $E_8\times E_8$ models, however, have a very different pattern of
anomaly cancellation, as described above.

Finally, in section 6, we shall discuss some phenomenological consequences of
our results. We start with some general remarks about possible applications to
model--building with anomalous $U(1)$ symmetries. We also point out that, within
special classes of compactifications, our results can be used to further
constrain the structure of the anomaly coefficients. As an example, we discuss
symmetric vacua which have recently been constructed~\cite{symm}. An important
issue related to anomalous $U(1)$ symmetries are Fayet--Illiopoulos terms. We
show that such terms arise from the dilaton part of the K\"ahler potential,
despite the different r\^ole of the dilaton. In accordance with the required
independence of the two sectors, we may have two FI terms, one in each sector.
They are proportional to the respective mixed $U(1)$ gauge anomaly coefficients.
Finally, we also analyze the gauge kinetic functions of the type I
$U(1)$ symmetries, which turn out to be more complicated than the standard ones
for the $\cH_n$ parts of the gauge group.

\vspace{0.4cm}

Before proceeding to the details of the M-theory anomaly cancellation, we
would like to raise the general question of whether the present results support the
idea that M-theory is simply $D=11$ supergravity, properly quantized (and thus
including its essential brane sectors). One indication that may go in this
direction is the analysis of M-theory/heterotic duality given in
Ref.~\cite{aspinwall}, taken in the context of $D=11$ supergravity with branes
wrapped and stacked around cycles of Calabi-Yau manifolds. In this
analysis, the Ho\v{r}ava-Witten orbifold $S^1/Z_2$ appears in a
certain singular limit of 11--dimensional supergravity on a $K3$
surface. The $E_8\times E_8$ gauge fields
on the orbifold fixed points arise from membranes wrapping spheres
within $K3$ that collapse in the singular limit. A central question here
is whether the full gauge group structure in the resulting $D=5$ theory can be
viewed as arising in a similar way. The results of
Ref.~\cite{aspinwall} would seem to suggest that this is indeed the
case for the analysis of sections three and five, that is, for all
five--dimensional theories obtained by a reduction of 11--dimensional
Ho\v{r}ava-Witten theory. The details of how that works will have to
be addressed in a future publication. 

%%%%%%%%%%%%%%%%%%%%%%%%%%%%%%%%%%%%%%%%%%%%%%%%%%%%%%%%%%%%%%%%%%%%%%%

\section{The five--dimensional effective action of heterotic
M--theory\label{sec:d5het}}

%%%%%%%%%%%%%%%%%%%%%%%%%%%%%%%%%%%%%%%%%%%%%%%%%%%%%%%%%%%%%%%%%%%%%%%

In this section we shall derive the five--dimensional effective action of
heterotic M--theory. This action has first been given in
Ref.~\cite{losw1,elpp,losw2} for the case of a standard embedding. Some of the
generalizations that emerge upon allowing general vector bundles in the vacuum
have been given in Ref.~\cite{nse}. However, certain aspects, particularly some
relevant for the question of anomaly cancellation, have not been explicitly
addressed in the literature. For the sake of clarity, we will therefore present
the reduction in a systematic way, focusing on issues related to anomaly
cancellation. This implies, in particular, that we need to consider the most
general structure for the $E_8$ vector bundles. 

\subsection{The Ho\v rava--Witten action in $D=11$}

The starting point of our reduction is the action for 11--dimensional
supergravity on an $S^1/Z_2$ orbifold, due to Ho\v rava and
Witten~\cite{hw1,hw2}. More precisely, we consider 11--dimensional
supergravity on the space $M_{11}=S^1/Z_2\times M_{10}$ where $M_{10}$
is a smooth 10--dimensional manifold. The orbifold coordinate is
denoted by $y$ throughout the paper and is taken to be in the range
$y\in [-\p\r ,\p\r]$. The $Z_2$ symmetry acts as $y\rightarrow -y$ (or $y
\rightarrow -y +2a$ for reflections around points other than $y=0$). Hence,
selecting the reflection point $y=0$ (with a conjugate reflection point at the
reidentification point $y=\pm\p\r$) there exist two 10--dimensional fixed
hyperplanes, $M_{10}^{(1)}$ at $y=y_1\equiv 0$ and $M_{10}^{(2)}$ at
$y=y_2\equiv\p\r$. The bulk action is given by 11--dimensional supergravity with
the metric $g$ and the three--index antisymmetric tensor field $C$ as the
bosonic fields. In the bulk, the field strength $G$ of $C$ is given by
$G=dC$. This bulk theory is coupled to two 10--dimensional $E_8$
super--Yang--Mills multiplets each residing on one of the orbifold fixed
planes. We denote the corresponding gauge fields by $A_n$ and their
field strengths by $F_n$, where $n=1,2$. Generators $T^a$ 
are always normalized such that $\tr (T^aT^b)=\d^{ab}$ where, as usual,
$\tr$ is defined to be $1/30$ of the trace in the adjoint.

The action for this theory can be organized as an expansion in powers
of $\k^{2/3}$, where $\k$ is the 11--dimensional Newton constant.
In order to simplify the notation, it is helpful to introduce the specific
combination~\footnote{In Ho\v{r}ava and Witten's original formulation
of the theory the coefficient $c$ was determined to be
$c=1$. Subsequently, it was found in Ref.~\cite{conrad, har} that
$c=2^{-1/3}$. In this paper, we will use $c=2^{-1/3}$.}

\begin{equation}
 \l = \frac{c}{2\sqrt{2}\p}\left(\frac{\k}{4\p}\right)^{2/3}\; .
 \label{lambda}
\end{equation}
Then the action has the structure
\begin{equation}
 S_{11}=S_0+S_1+S_2\ ,
\end{equation}
where the subscripts on the right hand side refer to the order in
$\l$. In the following, we shall focus on the bosonic part of this action,
which will be sufficient for the purpose of the present paper. To zeroth
order, we have the action of 11--dimensional supergravity
\begin{equation}
 2\k^2\, S_0=-\int_{M_{11}}\left[\sqrt{-g}\, R+G\w
             *G+\frac{\sqrt{2}}{3}  \, C\w G\w
 G\right]\; .\label{sugra}
\end{equation}
The Yang--Mills theories on the orbifold fixed planes appear at first
order in $\l$. They are specified by
\begin{equation}
 2\k^2\, S_1 = -\frac{\l}{\sqrt{2}}\sum_{n=1}^2\int_{M_{10}^{(n)}}
               \sqrt{-g}\left[\tr F_n^2-\frac{1}{2}\tr R^2\right]\; .
\end{equation}
At order $\l^2$, a Green--Schwarz term~\cite{dlm} and associated
$R^4$ terms~\cite{gg1,gg2} arise in the bulk. Here, we only need the
Green--Schwarz term which is given by
\begin{equation}
 2\k^2\, S_2 = \frac{\sqrt{2}}{3}\l^2\int_{M_{11}}C\w X_8\label{X8}
\end{equation}
where the anomaly polynomial $X_8$ reads~\footnote{Here and in the
following we will frequently omit the wedge symbol $\w$ in writing wedge
products in order to simplify the notation.}
\begin{equation}
 X_8 = -\frac{1}{8}\tr R^4+\frac{1}{32}(\tr R^2)^2\; .
\end{equation}
Finally, the Bianchi identity for $G$ has to be modified by source terms
supported on the orbifold planes that appear at order $\l$. This
modified Bianchi identity reads
\begin{equation}
 dG=-\l\sum_{n=1}^2J_n\w dy\,\d(y-y_n)\; .\label{bianchi}
\end{equation}
The sources $J_n$ depend on the gauge field and the curvature on the
orbifold planes and are explicitly given by
\begin{equation}
 J_n=\tr F_n^2-\frac{1}{2}\tr R^2\left.\right|_{y=y_n}\; .\label{J}
\end{equation}

\subsection{Vacuum configuration}

We would like to consider the above theory on a space--time $M_{11}$
with structure
\begin{equation}
 M_{11}=X\times S^1/Z_2\times M_4\label{M11}
\end{equation}
where $X$ is a Calabi--Yau three--fold. Eventually, we will be
interested in the five--dimensional theory on $M_5=S^1/Z_2\times M_4$
that is obtained by compactification on the Calabi--Yau space $X$.
Let us describe the background configurations adequate for such a
compactification~\cite{losw1,losw2}.  

In the bulk, we have a Calabi--Yau background
metric $\bar{g}$ and an associated curvature two--form $\bar{R}$
related to the tangent bundle $TX$ of the Calabi--Yau space $X$.
Here and in the following we use the bar to indicate fields with
components exclusively in the internal space $X$.

Furthermore, we should specify
the internal parts of the $E_8$ gauge fields. Rather than making any
specific choice, such as the standard embedding of the spin connection
into the gauge group, we would like to consider the general situation.
That is, we would like to cover all choices of internal gauge fields
compatible with the consistency requirements imposed by the theory.
Allowing for this most general situation is, of course, crucial in our
context since those gauge fields play a key role in anomaly
cancellation. We start with two $E_8$ vector bundles
$V_1$ and $V_2$ which are restricted so as to preserve supersymmetry, 
possessing thus a property that is known formally as semi-stability (see, \eg
\cite{dg}). Integrating the Bianchi identity~\eqref{bianchi} over
a five--cycle consisting of a four--cycle within the Calabi--Yau space
times the orbifold leads to the familiar constraint
\begin{equation}
 {\rm ch}_2(V_1)+{\rm ch}_2(V_2)={\rm ch}_2(TX)\label{ano1}
\end{equation}
on those bundles. Here ${\rm ch}_2$ is the second Chern
character. The field strengths associated to the bundles
$V_n$ are denoted by $\bar{F}_n$, where $n=1,2$. We also introduce
the internal parts $\bar{J}_n$ of the sources in the Bianchi
identity~\eqref{J} by
\begin{equation}
 \bar{J}_n=\tr\bar{F}_n^2-\frac{1}{2}\tr\bar{R}^2\left.\right|_{y=y_n}
           \; .\label{Jbar}
\end{equation}
The consistency relation~\eqref{ano1} can then be written in the
equivalent form
\begin{equation}
 \int_{C_{4i}}(\bar{J}_1+\bar{J}_2)=0
\end{equation}
where $\{C_{4i}\}_{i=1,\dots ,h^{1,1}}$ is a basis of Calabi--Yau
four--cycles. Furthermore, it is useful to introduce the topological
indices $\b_i$ and $\g_i$ by
\bea
 \b_i &=&-\frac{1}{8\p^2}\int_{C_{4i}}\bar{J}_1 \nn \\
      &=&-\frac{1}{8\p^2}\int_{C_{4i}}\left[\tr\bar{F}_1^2 
       -\frac{1}{2}\tr\bar{R}^2\right]\label{bi}\\
 \g_i &=& -\frac{1}{8\p^2}\int_{C_{4i}}\tr\bar{R}^2\; .\label{gi}
\eea

It is important to be somewhat more specific about the group structure
of our vector bundles. Following Ref.~\cite{dsww,dg} we write the structure
groups $\bcG_n$ of the bundles $V_n$ as products
\begin{equation}
 \bcG_n = \bcH_n\times\cJ_n\; .
\end{equation}
Here $\bcH_n$ is the semi--simple part and $\cJ_n$ contains the $U(1)$
factors. We also allow for Wilson lines which can be thought of as 
discrete parts of the gauge bundle. We choose the associated discrete structure
group to be part of $\cH_n$. Correspondingly, we write the bundles
and the gauge field strengths as
\begin{equation}
 V_n = W_n\oplus\bigoplus_a{\cal L}_n^a\; ,\qquad
 \bar{F}_n=\bar{f}_n+\sum_a\bcF_n^aQ_n^a\label{gd}
\end{equation}
where the bundles $W_n$ and associated gauge fields $\bar{f}_n$
correspond to the semi--simple part (and possible Wilson lines) 
while the line bundles ${\cal L}_n^a$ with $U(1)$ gauge fields $\bcF_n^a$
correspond to the various $U(1)$ factors. The associated $U(1)$
generators are denoted by $Q_n^a$. Here and in the following we label
these $U(1)$ factors by indices $a,b,\dots$. In order to satisfy the
field equations, the $U(1)$ field strengths have to be harmonic
two--forms. Hence they can be written as
\begin{equation}
 \bcF_n^a = v^{-1/3}\eta_{na}^i\O_i\; .\label{cFdef}
\end{equation}
Here $\{\O_i\}_{i=1,\dots ,h^{1,1}}$ is a basis of harmonic $(1,1)$
forms on the Calabi--Yau space dual to the basis of four--cycles
mentioned above. For dimensional reasons, we have also included a
power of the Calabi--Yau coordinate volume $v$ in the above definition. 
As usual, the coefficients $\eta_{na}^i$ in this
expansion are quantized in suitable units. We have mentioned above
that the vector bundles $V_n$ have to be semi--stable so as to
preserve supersymmetry. In particular, this implies for the line
bundles ${\cal L}_n^a$ that
\begin{equation}
 \O\w\O\w c_1({\cal L}_n^a)=0\label{cons1}
\end{equation}
where $\O$ is the K\"ahler form. Clearly, this condition cannot be satisfied
for $h^{1,1}=1$. Hence, only for Calabi--Yau spaces with
$h^{1,1}>1$ can the $U(1)$ factors introduced above occur.
The conditions~\eqref{cons1} in fact constitute restrictions on the
Calabi--Yau K\"ahler moduli space~\cite{dsww}. More explicitly, this
can be seen by writing the K\"ahler form as $\O=a^i\O_i$ with the
$(1,1)$ moduli $a^i$. Then, inserting this into Eq.~\eqref{cons1} along
with the expansion~\eqref{cFdef} of the $U(1)$ gauge fields leads to
\begin{equation}
 d_{ijk}\eta_{na}^ia^ja^k=0\; .\label{cons2}
\end{equation}
The constants $d_{ijk}$ are the Calabi--Yau triple intersection
numbers defined as
\begin{equation}
 d_{ijk} = \int_X\O_i\w\O_j\w\O_k\; .\label{d}
\end{equation}
In order to satisfy the equations~\eqref{cons2} we have to adjust
the K\"ahler moduli since the coefficients $\eta_{na}^i$, once chosen,
are fixed due to the quantization rule. 

\vspace{0.4cm}

Finally, we have to specify the background value of the antisymmetric
tensor field strength $G$. In fact, a non--vanishing internal $G$ is
forced upon us by the Bianchi identity~\eqref{bianchi} taking into
account the generically non--trivial internal sources $\bar{J}_n$.
This background value for $G$, also called a non--zero mode, plays an
important role in the derivation of the five--dimensional effective
theory, as has been demonstrated in Ref.~\cite{losw1,losw2}. As we
will see, taking into account this non--zero mode is also crucial in
order to understand anomaly cancellation in five dimensions.
Solving the Bianchi identity~\eqref{bianchi} subject to the
sources~\eqref{Jbar} along with the equation of motion for $G$ leads
to the following expression for the non--zero mode
\begin{equation}
 \bar{G} = 4\p^2\l\,\b_i\n^i\, \e (y)\; .\label{nonzero}
\end{equation}
Here $\{\n^i\}_{i=1,\dots ,h^{1,1}}$ forms a basis of harmonic
$(2,2)$ forms on the Calabi--Yau space dual to the basis $\{\O_i\}$ of
harmonic $(1,1)$ forms. The step--function $\e (y)$ is defined to be
$+1$ for $y\geq 0$ and $-1$ otherwise. 

Note that the presence of the step--function in this background value for $G$
breaks the five-dimensional translation invariance at the selected orbifold points
$y=0$, $y=\pm\p\r$. Accordingly, the $D=5$ field theory will be found to have
extended--objects at the locations of the step--function jumps.  At the present
time, however, we see only a hint of this structure in \eqref{nonzero}, which
constitutes a generalization of the usual kind of Kaluza-Klein ansatz for field
configurations in the internal dimensions. Note finally that although
\eqref{nonzero} breaks the $D=5$ translation invariance, it actually {\em
restores} the $Z_2$ symmetry $y\rightarrow -y$ of the $D=11$ theory, which would
otherwise be broken if the step--function were not present in 
\eqref{nonzero}, since $G$ is a $Z_2$ odd quantity.

\subsection{Zero modes}

We would now like to summarize the structure of zero modes arising for
backgrounds of the above type. Let us start with bulk zero modes.
As we will see, for the discussion of five--dimensional anomalies, we
can focus on the zero modes of the antisymmetric tensor field related
to the constant Calabi--Yau mode and the $(1,1)$ sector. In addition,
we will consider the bulk modes needed to complete $\cN=1$ multiplets
in five dimensions.

The Calabi--Yau breathing mode $V$ is defined as
\begin{equation}
 V=\frac{1}{v}\int_X\sqrt{\bar{g}}\; .
\end{equation}
In the previous subsection we have already introduced the $(1,1)$
moduli $a^i$, where $i=1,\dots ,h^{1,1}$. Those are, however, not
independent from the volume modulus $V$. To remove the redundancy, it
is useful to define the ``shape moduli'' $b^i$ by
\begin{equation}
 b^i=V^{1/3}a^i\; .
\end{equation}
Those $h^{1,1}$ quantities are subject to one constraint. Hence they
represent only $h^{1,1}-1$ independent degrees of freedom.
Neglecting contributions from harmonic $(2,1)$ forms, the antisymmetric
tensor three--form and its field strength can be written as
\bea
 C &=& \bar{C}+\tilde{C}+\x\O_3+\x^*\O_3^*+
       \cB^i\w\O_i \\
 G &=& \bar{G}+\tilde{G}+X\w\O_3+X^*\w\O_3^*+\cD^i\w\O_i\; .
\eea
Here $\bar{C}$ and $\bar{G}$ represent the non--zero mode background
specified in the previous subsection. The field $\tilde{C}$, with field
strength $\tilde{G}$, represent a three--index antisymmetric tensor field in
the external five dimensions that is associated to the constant mode on the
Calabi--Yau space. The complex scalar $\x$ and its field strength $X$ arise
from the harmonic $(3,0)$ form $\O_3$. Furthermore, we have $h^{1,1}$ gauge
fields $\cB^i$ with field strengths $\cD^i$ related to the harmonic $(1,1)$
forms $\O_i$.

In five dimensions, these zero modes give rise to the following
multiplets. The five--dimensional metric along with a certain linear
combination of the vectors $\cB^i$ form the bosonic part of the
$\cN =1$ gravity multiplet. The remaining $h^{1,1}-1$ vectors $\cB^i$
together with the scalars $b^i$ represent $h^{1,1}-1$ vector
multiplets. Finally, the volume modulus $V$, the complex scalar $\x$
and the dual of the three--form $\tilde{C}$ (which is a scalar in five
dimensions) form the universal hypermultiplet.

\vspace{0.4cm}

Next we should consider the zero modes arising on the orbifold planes.
After compactification, these planes become four--dimensional
hyperplanes $M_4^{(n)}$ located at $y=y_n$ in the five--dimensional
space. Recall, that the internal gauge groups have a product structure
$\bcG_n=\bcH_n\times\cJ_n$ where $\bcH_n$ are the semi--simple parts
and $\cJ_n$ contains the $U(1)$ factors. The surviving
low--energy groups $\cG_n$ are the commutants of these internal
groups within $E_8$. Their structure is given by
\begin{equation}
 \cG_n = \cH_n\times\cJ_n\; .
\end{equation}
where $n=1,2$. Note that the $U(1)$ factors commute with themselves.
Hence $\cJ_n$ is part of the low--energy gauge groups as well. We will call
such $U(1)$ factors in $\cJ_n$ that have a counterpart in the internal
structure group $U(1)$ symmetries of type I, or $U_I(1)$ in short.
The remaining parts of the low--energy groups contain the non--abelian
factors and are denoted by $\cH_n$. Note, however, that the
groups $\cH_n$ do not necessarily have to be semi--simple. The
low--energy group may contain $U(1)$ factors other than the type I ones
mentioned above, which would then be contained in $\cH_n$. We will call
these type II $U(1)$ symmetries, or $U_{II}(1)$ in short. It is clear, from
their relation to the internal structure groups, that type I and type II
$U(1)$ fields are of very different nature. This will become apparent in the
discussion of anomaly cancellations. We also introduce gauge fields
$\tilde{A}_n$ with field strengths $\tilde{F}_n$ for the group $\cH_n$ and
$U(1)$ gauge fields
$\tcA_n^a$ with field strengths $\tcF_n^a$ for the type I $U(1)$ factors in
$\cJ_n$. 

To discuss anomalies, we also need some information about the $\cN =1$
chiral multiplets on the orbifold planes. As usual we decompose the
adjoint of $E_8$ under the subgroup $\bcG_n\times\cG_n$ as
\begin{equation}
{\bf 248}_{E_8}\rightarrow \bigoplus_r (\bar{L}_n^r,L_n^r)\label{decomp}
\end{equation}
where the sum runs over all representations $(\bar{L}_n^r,L_n^r)$ of 
$\bcG_n\times\cG_n$. Then, in general, we expect chiral $\cN =1$
multiplets in all representation $L_n^r$ of the external gauge group
$\cG_n$ that appear in this decomposition. Fortunately, all we need to
know in this context is the chiral asymmetry of such multiplets in
$L_n^r$ which we denote by $N_n^r$. From the index theorem this
asymmetry is given by
\begin{equation}
 N_n^r = \frac{1}{6}\frac{1}{(2\p )^3}\int_X\left[\tr_{\bar{L}_n^r}\bar{F}_n^3
         -\frac{1}{8}\tr_{\bar{L}_n^r}\bar{F}_n\tr\bar{R}^2\right]\; .
 \label{index}
\end{equation}
where $\tr_{\bar{L}_n^r}$ denotes the trace taken in the representation
$\bar{L}_n^r$ of the internal gauge group $\bcG_n$. Recall also that
$\bar{F}_n$ and $\bar{R}$ are the internal gauge field and curvature
backgrounds. 

\subsection{The five--dimensional effective action}

We are now ready to summarize the parts of the five--dimensional
effective action that are essential for our purpose. Starting from
the 11--dimensional theory and using the backgrounds and the zero
modes introduced above one finds
\begin{equation}
 S_5=S_{\rm kin}+S_{\rm top}+S_{\rm bound}\; .\label{S5}
\end{equation}
Here $S_{\rm kin}$ contains the kinetic terms of the antisymmetric
tensor fields and is given by
\begin{equation}
 2\k_5^2\,S_{\rm kin}=-\int_{M_5}\left[2G_{ij}\cD^i\w
 *\cD^j+2V^{-1}X\w *X^*+V^2G\w *G\right]\; .\label{Skin}
\end{equation}
We have omitted the kinetic terms of all other moduli since they will
not be relevant for the subsequent discussion. The K\"ahler moduli
space metric $G_{ij}$ will not be explicitly needed here.
The topological part of the action reads
\begin{multline}
 2\k_5^2\,S_{\rm top} =
  -\sqrt{2}\int_{M_5}\left[\frac{\p^2\l^2}{6v^{2/3}}\;\g_i\,\cB^i\w\tr R^2
  -\frac{8\p^2\l}{v^{2/3}}\e (y)\;\b_i\,\cB^i\w G +\frac{1}{3}d_{ijk}
  \cB^i\cD^j\cD^k\right.\\
  \left. +i(\x G\w X^*-\x^*G\w X)\right]\; .
  \label{Stop}
\end{multline}
For the boundary part we have
\begin{equation}
 S_{\rm bound} =
  -\frac{1}{4g_0^2}\sum_{n=1}^2\int_{M_4^{(n)}}\sqrt{-g}\left[V\tr F_n^2
  +V\sum_a ({\cal F}_n^a)^2+(\mbox{matter})\right]\; .\label{Sbound}
\end{equation}
Furthermore, the above action has to be supplemented with the
following Bianchi identities
\bea
 dG &=& -\l\sum_{n=1}^2J_n\w dy\,\d(y-y_n) \label{b1}\\
 dX &=& (\mbox{matter})\\
 d\cD^i &=& -2v^{-1/3}\l\sum_{n=1}^2\eta_{na}^i\cF_n^a\w dy\,\d(y-y_n)
            +(\mbox{matter}) \label{b3}
\eea
where the sources are given by
\begin{equation}
 J_n = \tr F_n^2+\sum_a\cF_n^a\w\cF_n^a-\frac{1}{2}\tr R^2\; .
 \label{Jn}
\end{equation}
Those Bianchi identities directly descend from the 11--dimensional
one, Eq.~\eqref{bianchi}. It is also useful to write them in their
integrated form
\bea
 G &=& dC-\l\sum_{n=1}^2w_n\w dy\d(y-y_n) \label{bi1}\\
 X &=& d\x +(\mbox{matter}) \\
 \cD^i &=& d\cB^i-2v^{-1/3}\l\sum_{n=1}^2\eta_{na}^i\cA_n^a\w dy\d(y-y_n)
            +(\mbox{matter}) \label{bi3}
\eea
with the Chern--Simons form $w_n$ defined by
\begin{equation}
 dw_n=J_n\; .
\end{equation}
We have also introduced the five--dimensional Newton constant $\k_5$
and the gauge coupling $g_0$. They are given in terms of 11--dimensional
quantities by
\begin{equation}
 \k_5^2=\frac{\k^2}{v}\; ,\qquad g_0^2=\frac{\k_5^2}{\sqrt{2}\l}\; .
\end{equation}

A few comments concerning notation are in order. In the previous
subsections we have distinguished five--dimensional fields from their
11--dimensional counterparts by a tilde. From now on, we will be
working in five dimensions and we shall omit the tilde for notational
simplicity. So, for example, $C$ is the five--dimensional three--index
antisymmetric tensor field with field strength $G$. Furthermore, $A_n$
are the gauge fields with gauge group $\cH_n$ and field strength
$F_n$, while $\cF_n^a$ are the $U_I(1)$ vector fields with
associated gauge group $\cJ_n$.
We also recall that $d_{ijk}$ are the Calabi--Yau intersection
numbers~\eqref{d} and $\b_i$ and $\g_i$ are topological numbers
defined in terms of the
internal bundles by Eq.~\eqref{bi} and \eqref{gi}. Although matter
fields on the orbifold planes do play an important role in the
following their explicit contributions to the action and the Bianchi
identities is not essential. Their presence has, however, been indicated
in the above action.

It is instructive, for the following, to understand the 11--dimensional
origin of some of the above terms. For example, the first term in the
topological part of the action is the reduction of the $C\w X_8$
term~\cite{fkm} given in Eq.~\eqref{X8}. All other topological terms
originate from the Chern--Simons term $CGG$ of 11--dimensional
supergravity. Particularly, the second term in Eq.~\eqref{Stop} results by
taking one of the field strengths $G$ in $CGG$ to be the internal non--zero
mode~\eqref{nonzero}. This term causes the gauging of the universal
hypermultiplet as has been shown in Ref.~\cite{losw1,losw2}. It will also
be essential to understand anomaly cancellation in the five--dimensional
theory. Another essential part in the above action is the right hand side
of the Bianchi identity~\eqref{bi3}. The $U_I(1)$ gauge field strengths
$\cF_n^a$ in this Bianchi identity arise because their associated gauge
group
$\cJ_n$ is part of the internal structure group at the same time. In fact,
only in this case can one have a non-vanishing result for the trace $\tr
F^2$, where one $F$ is taken to have internal indices and the other one to
have external indices. It is for this reason, that the
$U_I(1)$ fields of type I in $\cJ_n$ play a special role. Note that the
$U_{II}(1)$ fields, contained in the $\cH_n$ part of the low-energy gauge
group, do not appear in the Bianchi identity~\eqref{bi3}.

%%%%%%%%%%%%%%%%%%%%%%%%%%%%%%%%%%%%%%%%%%%%%%%%%%%%%%%%%%%%%%%%%%%%%%%%%%%%%

\section{Anomaly cancellation in five dimensions}

%%%%%%%%%%%%%%%%%%%%%%%%%%%%%%%%%%%%%%%%%%%%%%%%%%%%%%%%%%%%%%%%%%%%%%%%%%%%%

We now shall study anomaly cancellation in the five--dimensional action of
heterotic M--theory that we have just derived. We expect the anomaly
cancellation to rely on an interplay between bulk and orbifold planes in
much the same way as in 11 dimensions. More precisely, in the
five--dimensional theory, the gauge theories on the orbifold planes might
have quantum anomalies at one loop due to the familiar triangle diagrams.
These apparent anomalies should then be cancelled by a classical gauge
anomaly of the bulk theory supported on those orbifold planes or, in other
words, by anomaly inflow from the bulk. Let us work this out in some detail,
starting with the anomalous variation of the bulk action.

\subsection{Classical variation of the bulk action}

An anomalous variation of the bulk action~\eqref{Skin}, \eqref{Stop}
should be triggered by the Bianchi identities \eqref{b1}--\eqref{b3}
as they represent the only way in which the orbifold gauge fields directly
communicate with the bulk fields. Clearly, then, the bulk
antisymmetric tensor fields are the only bulk fields that potentially
transform under the gauge transformations associated to the orbifold gauge
fields. As in 11 dimensions, our strategy will be to keep the
antisymmetric tensor field strengths invariant so that the kinetic
part~\eqref{Skin} of the action remains inert. This, however,
typically implies assigning non--trivial gauge transformations to the
antisymmetric tensor fields themselves. In the case at hand, we can
see from the  Eqs.~\eqref{bi1}--\eqref{bi3} that we need to assign
such transformation to the bulk three--form $C$ and the bulk vector
fields $\cB^i$. Specifically, while $C$ transforms under the
$\cH_n$ parts of the orbifold gauge groups, the vector fields
$\cB^i$ transform under the $U_I(1)$ gauge transformations of type I. We note,
however, that the bulk action depends on $C$ only through its field
strength $G$. Therefore, the variation of $C$ is not relevant for our
purpose. Hence, we can already draw the important conclusion that
the $\cH_n$ parts of the gauge groups have to be
non--anomalous. More precisely, the one--loop anomalies due to
$\cH_n$ triangle diagrams have to vanish since, as we have just
seen, there is no corresponding bulk variation available to cancel them.
We will come back to this in a more systematic way later on.

For now, we should be more explicit about the gauge variation of the
$U_I(1)$ fields and the transformation of the bulk vector fields that they
induce. We denote the transformation parameters of these $U_I(1)$ fields
$\L_n^a$ and write
\begin{equation}
 \d{\cal A}_n^a=d \L_n^a\; .
\end{equation}
Then, demanding that the vector field strengths $\cD^i$ be invariant, we
conclude from Eq.~\eqref{bi3} that the vector fields $\cB^i$
should transform as
\begin{equation}
 \d\cB^i=2v^{-1/3}\l\sum_{n=1}^2\eta_{na}^i\L_n^a\, dy\,\d (y-y_n)\; .
 \label{deltaB}
\end{equation}
As we have already mentioned, the kinetic part of the bulk
action~\eqref{Skin} remains invariant under this transformation. The
variation of the topological part~\eqref{Stop}, however, is
non--trivial and consists of terms that are supported purely on the orbifold
planes. The latter, of course, is caused by the delta functions in the
transformation law~\eqref{deltaB}. To further evaluate this anomalous
variation, we need to know the values of the antisymmetric tensor field
strengths $G$ and $\cD^i$ on the orbifold planes
$y=y_n$. To be more specific, we need only the components
of those fields transverse to the orbifold since that is all that the
anomalous variation depends on. The reason for this is the presence of
the $dy$ differential in Eq.~\eqref{deltaB} which projects onto those
components. Fortunately, those components are $Z_2$ odd and, hence,
their behavior close to the orbifold planes is completely determined
by the source terms in the Bianchi identities. Solving the Bianchi
identities~\eqref{b1} and \eqref{b3} along with the equations of motion
for $C$ and $\cB_n^a$ leads to
\bea
 dy\w\cD^i\left.\right|_{y=y_n} &=& \mp\l v^{-1/3}\eta_{na}^i\,\cF_n^a\w
   dy \\
 dy\w G\left.\right|_{y=y_n} &=& \mp\frac{\l}{2}\, J_n\w dy\; .
\eea
With these expressions we finally find for the anomalous variation of
the bulk action
\begin{multline}
 \d_{\rm cl}S_5 = -\frac{c^3}{128\p^3}\sum_{n=1}^2\int_{M_4^{(n)}}
                  \L_n^a\left[ (\mp 2\b_i\eta_{na}^i)\tr F_n^2
                  -\left(\mp\b_i-\frac{1}{12}\g_i\right)\eta_{na}^i\tr
                  R^2\right.\\
                  \left.+\left(\mp 2\b_i\eta_{na}^i\d_{bc}+\frac{1}
                  {6\p^2}d_{ijk}\eta_{na}^i\eta_{nb}^j\eta_{nc}^k
                  \right)\cF_n^b\cF_n^c\right]\; .
 \label{dS5}
\end{multline}
Here and in the following, when we use alternating signs, the upper sign
refers to the first orbifold plane, $n=1$, while the lower sign refers to
the second plane, $n=2$. The above classical variation of the bulk action
should be cancelled by quantum anomalies on the orbifold planes that
originate from the usual triangle diagrams. That this is possible at all is
due to the obvious but nonetheless important fact that the gauge variation
of the bulk action consists solely of terms supported on the orbifold
planes.  From the form of the above variation it is obvious which types of
triangle diagram should be responsible for the cancellation. The first term
in~\eqref{dS5} corresponds to a mixed gauge anomaly with one $U_I(1)$ current
of type I and two currents from the $\cH_n$ parts of the gauge groups. The
second term should be cancelled by a mixed gravitational anomaly with one
$U_I(1)$ current and two gravity currents. Finally, the structure
of the third term corresponds to a cubic anomaly of the $U_I(1)$ gauge
fields. In the following subsection we will show explicitly that the
anticipated cancellation indeed works.

\subsection{Quantum variation on the orbifold planes}

Above we have mentioned the types of triangle diagrams whose anomalous
variation should be cancelled by~\eqref{dS5}. Clearly, since we have done
our reduction from $D=11$ to $D=5$ consistently, we expect this
cancellation to work. The consistently derived five--dimensional
effective theory should be anomaly-free in the same way as the 11
dimensional action. We still find it useful, if only as a
cross--check of our result, to verify this explicitly. To do this, we
need to know the anomaly coefficients of the various types of triangle
diagrams defined as follows
\begin{subeqnarray}
 \cC_n &=& \sum_rN_n^r\,\tr_{L_n^r}(T_n^3)  \label{cd1} \\
 \cC_{na} &=& \sum_rN_n^r\,\tr_{L_n^r}(Q_n^aT_n^2)  \label{cd2} \\
 \cC_{nab} &=& \sum_rN_n^r\,\tr_{L_n^r}(Q_n^aQ_n^bT_n) \label{cd3} \\
 \cC_{nabc} &=& \sum_rN_n^r\,\tr_{L_n^r}(Q_n^aQ_n^bQ_n^c) \label{cd4} \\
 \cC_n^{(L)} &=& \sum_rN_n^r\,\tr_{L_n^r}(T_n) \label{cd5} \\
 \cC^{(L)}_{na} &=& \sum_rN_n^r\,\tr_{L_n^r}(Q_n^a)\; .\label{cd6}
\end{subeqnarray}
Here $T_n$ denotes any generator in the $\cH_n$ parts of the gauge group. Recall
that $Q_n^a$ are the generators of the $U_I(1)$ gauge groups. Furthermore,
$L_n^r$ denotes the possible matter representations of the low energy
gauge group $\cG_n$
on the orbifold plane $n$, that is, the representations appearing in the
decomposition~\eqref{decomp}. Fortunately, all that enters the anomaly is the
chiral asymmetry $N_n^r$ of those representations which can be expressed in terms
of the internal bundles via the index theorem~\eqref{index}. The first four
anomaly coefficients above cover the possible combinations of $U_I(1)$ gauge
fields with gauge fields in $\cH_n$. The final two coefficients measure the mixed
gravitational anomalies of $\cH_n$ and the $U_I(1)$ fields, respectively.
Note that although there are no purely gravitational anomalies
in five dimensions, the mixed gravitational-gauge anomalies involve
the orbifold planes, which are
four-dimensional. In four dimensions, mixed gravitational--gauge anomalies can
arise \cite{mixedgauge,agw}, which is precisely what happens in
(\ref{cd5}e,\ref{cd6}f).

Now, following Ref.~\cite{w1,gsw2}, we replace $N_n^r$ in the above
anomaly coefficients using the index theorem and further simplify the
expressions by means of the trace formulae collected in Appendix A.
This allows one to completely express the anomaly
coefficient in terms of topological Calabi--Yau and bundle data. We find
\begin{subeqnarray}
 \cC_n &=& 0 \label{c1} \\
 \cC_{na} &=& \mp\frac{1}{8\p}\eta_{na}^i\b_i \label{c2} \\
 \cC_{nab} &=& 0 \label{c3} \\
 \cC_{nabc} &=& \frac{3}{8\p}\left[\mp \b_i\eta_{n(a}^i\d_{bc)}+
                \frac{1}{12\p^2}d_{ijk}\eta_{na}^i\eta_{nb}^j
                \eta_{nc}^k\right] \label{c4} \\
 \cC^{(L)}_n &=& 0 \label{c5} \\
 \cC^{(L)}_{na} &=& \frac{3}{2\p}\left(\mp\b_i-\frac{1}{12}\g_i\right)
                    \eta_{na}^i\; .\label{c6}
\end{subeqnarray}
As before, the upper (lower) sign refers to the first (second)
orbifold plane. The quantum variation due to the triangle diagrams is
then given by
\begin{equation}
 \d_{\rm Q}S_5 = \frac{1}{16\p^2}\sum_{n=1}^2\int_{M_4^{(n)}}\L_n^a\left[
                 \cC_{na}\tr F_n^2-\frac{1}{24}\cC_{na}^{(L)}\tr R^2
                 +\frac{1}{3}\cC_{nabc}\cF_n^b\cF_n^c\right]\label{dQS5}
\end{equation}
with the anomaly coefficients $\cC$ as given in Eqs.~(\ref{c1}a--\ref{c6}f).
This quantum variation cancels the classical one in Eq.~\eqref{dS5}
except for the first term in the cubic anomaly coefficient.
In the quantum variation~\eqref{dQS5}, this first term is symmetrized
in all three indices while this is not the case in the classical
variation~\eqref{dS5}. This, however, is not a problem. The form of
the anomaly in Eq.~\eqref{dQS5} that we have chosen corresponds to a
specific way of regularizing the triangle diagrams. Using the known
ambiguity in the four--dimensional anomaly, we can, however,
regularize the diagrams associated to the cubic anomaly differently
and put them exactly in the appropriate form so as to cancel the
classical variation~\eqref{dS5}. Hence, we indeed have
$(\d_{\rm cl}+\d_{\rm Q})S_5=0$ and the total anomaly cancels.

\subsection{Discussion}

Let us now discuss the significance of the above results in some
detail. First, we would like to do this in relation to the
11--dimensional origin of the five--dimensional theory.

As we have seen, the anomalous classical variation originates from the
first three terms in the topological part~\eqref{Stop} of the
five--dimensional bulk action. The 11--dimensional origin of these
terms is as follows
\bea
 C\w X^8&\rightarrow&\g_i\,\cB^i\tr R^2 \label{o1} \\
 CGG \mbox{ with non-zero mode}&\rightarrow& \b_i\,\cB^i\w G \label{o2}\\
 CGG&\rightarrow&d_{ijk}\cB^i\cD^j\cD^k\; .\label{o3}
\eea
Recall here that $CGG$ is the Chern--Simons term of 11--dimensional
supergravity~\eqref{sugra} while $C\w X_8$ is the Green--Schwarz term
defined in Eq.~\eqref{X8}. The non--zero mode that, inserted into
$CGG$, leads to the second term above, is a purely internal
configuration for the 11--dimensional antisymmetric tensor fields
strength $G$. It was explicitly given in Eq.~\eqref{nonzero}. 
The crucial ingredient forcing us to assign gauge transformations to
the bulk antisymmetric tensor fields $\cB^i$ was the Bianchi
identity~\eqref{bi3}. It directly results from the 11--dimensional
Bianchi identity~\eqref{bianchi} with two of the indices being taken
internal and the others external. As we have seen, given such a
configuration of indices, the sources in the five--dimensional Bianchi
identities for $\cD^i$ can only receive non--trivial contributions
from the $U_I(1)$ gauge fields. As a consequence, only those
$U_I(1)$ fields appear on the right hand side of~\eqref{bi3} and
the bulk vector fields $\cB^i$ transform under the type I gauge
transformation only. What is the individual contribution of the above
three bulk terms to the anomalous variation? The first
term~\eqref{o2}, originating from the 11--dimensional Green--Schwarz
term, causes only the second contribution proportional to $\g_i$ in the
mixed gravitational anomaly in Eq.~\eqref{dS5}. Likewise, the
third term~\eqref{o3}, originating from the 11--dimensional
Chern--Simons term, leads to only one term, namely the second one in
the cubic anomaly in Eq.~\eqref{dS5}. The most important part of the
anomalous variation results from the second term~\eqref{o2}. It leads
to all terms in Eq.~\eqref{dS5} proportional to $\b_i$ and,
hence, contributes to all three types of anomalies that we have
encountered. This shows again the importance of this term, which is
also responsible for the gauging of the supergravity
theory~\cite{losw1,losw2}. Had we missed this term, or equivalently,
had we missed the underlying non--zero mode, we would clearly be unable to
uncover the five--dimensional anomaly structure.

\vspace{0.4cm}

Let us now move on to discuss this five--dimensional structure in
detail. To do this, let us focus on the expression~\eqref{dQS5} for
the quantum variation. The non--trivial information resides in the
comparison between the general definition (\ref{cd1}a--\ref{cd6}f)
of the anomaly coefficients and the expressions (\ref{c1}a--\ref{c6}f)
that one obtains for those coefficients within the context of
heterotic M--theory. A priori, from their
definition (\ref{cd1}a--\ref{cd6}f), the anomaly coefficients could
have taken any value depending on the gauge groups and particle
contents on the orbifold planes. We are told, however, that in
heterotic M--theory they have a precise structure in terms of the
underlying compactification. More
concretely, they are given by the Calabi--Yau intersection numbers
$d_{ijk}$, the topological numbers $\b_i$ and $\g_i$ related to second
Chern characters of the internal gauge and tangent bundles and the
coefficients $\eta_{na}^i$ that specify the $U(1)$ parts of the
internal gauge bundles. Clearly, this structure reflects the
restrictions that heterotic M--theory imposes on the particle content
residing on the orbifold planes. In a ``bottom up approach'' we can
also turn this around and use the anomaly structure to learn about
these restrictions. For a more precise discussion, we first remark that the
expressions (\ref{c1}a--\ref{c6}f) are rather symmetric with respect
to the two orbifold planes. It is worth pointing out that
the two gauge theories on those planes are completely hidden with
respect to one another, that is, they only interact gravitationally.
Correspondingly, anomaly cancellation works completely independently for
the two planes. Obviously, there cannot be any $U(1)$ gauge fields
``extending over both planes'' not to mention $U(1)$ gauge fields that
``got displaced into the bulk in between''. Consequently, we have only
non--vanishing triangle diagrams that couple gauge currents on the
{\em same} plane. Our subsequent discussion, focusing
on a single orbifold plane, therefore applies to either one of the two
orbifold planes.

The first crucial observation is that some of the anomaly coefficients
vanish within heterotic M--theory. This implies that the classical as well
as the quantum variation should be zero individually. Specifically, we see
from Eq.~(\ref{c1}a) and (\ref{c5}e) that the diagrams coupling three currents
in the $\cH_n$ part of the gauge group as well as the mixed gravitational
anomaly with one $\cH_n$ and two gravity currents vanish. In this sense,
the $\cH_n$ part of the gauge group is anomaly--free. Recall that $\cH_n$
contains the complete non--abelian part of the gauge group as well as the
$U_{II}(1)$ fields of type II. It is well--known that the non--abelian part
of the gauge group is anomaly--free~\cite{w1}. In addition, we learn here
that the same is true for $U_{II}(1)$ factors of type II. As a consequence,
non--vanishing diagrams always involve the $U_{I}(1)$ fields of type I.
Among those diagrams the one coupling two $U_{I}(1)$ fields to a field in
$\cH_n$ vanishes from Eq.~(\ref{c3}c). This is trivial for the non--abelian
part of $\cH_n$ (since then $\tr_{L_n^r}(T_n)=0$ always). However, it gives
a non--trivial information for the $U_{II}(1)$ factors in $\cH_n$, namely
that the $U_{I}(1)^2U_{II}(1)$ triangle diagram vanish. We are left with
three types of anomaly that can be non--zero in general. These are the mixed
$U_{I}(1)$ gauge anomaly (\ref{c2}b), the $U_{I}(1)^3$ cubic
anomaly (\ref{c4}d) and the mixed $U_{I}(1)$ gravitational
anomaly (\ref{c6}f). From the general form of the mixed $U_{I}(1)$ gauge
anomaly, there could in principle be a dependence on the specific generator
$T_n$ of $\cH_n$ that has been chosen. However, Eq.~(\ref{c2}b) shows that
the anomaly coefficient is, in fact, independent of this choice. Hence, we
conclude that the mixed $U_{I}(1)$ gauge anomaly coefficient is the same
for all non--abelian factors and all $U_{II}(1)$ factors in $\cH_n$. In
this sense, the mixed gauge anomaly may be called universal.
 
How many anomalous $U_I(1)$ symmetries are there? As we have seen, we start
with a given number $k_n$ of $U_I(1)$ factors on each orbifold plane
$n=1,2$ which is determined by the specific compactification. Not all of
them, however, necessarily have to be anomalous. It is clear from
Eq.~(\ref{c2}b), for example, that we can always choose a basis such that
only one of the $U_I(1)$ factors has a mixed gauge anomaly while this type
of anomaly vanishes for all the other
$U_I(1)$. Here we stress again that this statement as well as the
subsequent ones apply to each orbifold plane separately. Hence, in sum, we
may have two $U_I(1)$ groups with mixed gauge anomalies, one on each
orbifold plane.  Each of these $U_I(1)$'s may have mixed gravitational and
cubic anomalies at the same time. Those are fixed in terms of topological
data by Eq.~(\ref{c4}d) and (\ref{c6}f). However, these expressions also show
that the three anomaly coefficients are not related in any universal way.
That is to say, without further information about the type of
compactification and the associated topological data, they are practically
independent. We have therefore identified one $U_I(1)$ factor on each
orbifold plane that has a universal mixed gauge anomaly (independent of the
gauge factor) as well as generally unrelated cubic and mixed gravitational
anomalies. While the remaining $k_n-1$ $U_I(1)$ factors are now free of
mixed gauge anomalies by construction, they may have gravitational and cubic
anomalies, however. Again, we can diagonalize these remaining
$U_I(1)$'s such that at most one of them has a mixed gravitational and a
cubic anomaly. Finally, then, the other $k_n-2$ $U_I(1)$'s which are now
free of mixed gauge and gravitational anomalies may have cubic anomalies
only.

Let us summarize this discussion briefly. Recall that $\cH_n$ consist of
the non--abelian and the $U_{II}(1)$ gauge fields, collectively called
$A_n$. Furthermore, we denote $U_I(1)$ gauge fields by
$\cA_n^a$, where $a=1,\dots ,k_n$ and the graviton by $g$. We then refer to
a triangle diagram by specifying the triple of gauge fields to which it
couples. Then the anomaly structure on {\em each} orbifold plane in
five--dimensional heterotic M--theory is as follows~:
\begin{itemize}
 \item The $\cH_n$ part of the gauge group is anomaly--free in the
        sense that the $A_n^3$ and $A_ngg$ anomalies vanish.
 \item The mixed $\cA_n^a\cA_n^bA_n$ anomalies vanish.
 \item After a suitable choice of basis there is at most a single $U_I(1)$
       gauge field, say $\cA_n^1$ with all three remaining types of
       anomalies non--vanishing. That is, we may have anomalies of
       type $\cA_n^1A_nA_n$, $\cA_n^1\cA_n^a\cA_n^b$ and $\cA_n^1gg$. In terms
       of the topological data the associated anomaly coefficients are
       given in Eq.~(\ref{c2}b), (\ref{c4}d) and (\ref{c6}f).
       The remaining $U_I(1)$ factors are free of mixed gauge
       anomalies, that is the $\cA_n^aA_nA_n$ anomaly vanishes for $a>1$. 
 \item The coefficient of $\cA_n^1A_nA_n$ is independent of which
       specific gauge field $A_n$ within $\cH_n$ is considered. Apart
       from this restriction, the three non--vanishing anomaly coefficients are
       generically unrelated.
 \item After another choice of basis there is at most one among the
       remaining $U_I(1)$ symmetries, say $\cA_n^2$, with
       mixed gravitational and cubic anomaly. In other words, the anomalies
       of type $\cA_n^2gg$ and $\cA_n^2\cA_n^a\cA_n^b$ can be
       non--vanishing. Again the two associated anomaly coefficients
       are generically unrelated. All other $\cA_n^a$ for $a>2$ have cubic
       anomalies at most.
\end{itemize}

%%%%%%%%%%%%%%%%%%%%%%%%%%%%%%%%%%%%%%%%%%%%%%%%%%%%%%%%%%%%%%%%%%%%%%%%%%%%%

\section{Five--dimensional anomaly cancellation -- a different
        viewpoint}

%%%%%%%%%%%%%%%%%%%%%%%%%%%%%%%%%%%%%%%%%%%%%%%%%%%%%%%%%%%%%%%%%%%%%%%%%%%%%

In the previous section, we have analyzed anomaly constraints on the
particle spectrum residing on the orbifold planes of five--dimensional
heterotic M--theory. In doing so we have used all of the
knowledge arising from the 11--dimensional Ho\v rava--Witten
construction upon descent down to five dimensions. The derivation of this
11--dimensional theory and, in particular, the $E_8$ gauge multiplets
to which it couples is, however, based upon anomaly cancellation as well. 
A natural question, therefore, is whether one can avoid using some of this
11--dimensional knowledge and derive the anomaly constraints directly on a
five--dimensional basis. Concretely, we would like to start from pure
11--dimensional supergravity, that is, we are not coupling any $E_8$
gauge fields to the theory. We can then consider this theory in the
background of a Calabi--Yau three--fold and a non--zero mode
configuration of the antisymmetric tensor field. The resulting
effective action is a five--dimensional $\cN =1$ gauged supergravity,
given by the bulk part of the action that we have described in
the previous section. Subsequently, we would like to consider this
five--dimensional theory on the orbifold $S^1/Z_2$. Our main question
is then the following. Which constraints does five--dimensional
anomaly cancellation impose on the ``twisted states'' residing on
the two four--dimensional fixed planes of this orbifold? Hence, we are
asking, in analogy with the original 11--dimensional construction,
how five--dimensional M--theory can be put on an orbifold
consistently. Given that one is usually bound to use anomaly cancellation as
the basic tool, this question might be no less fundamental than the
corresponding one in 11 dimensions.

\subsection{Starting point -- five--dimensional gauged supergravity}

We start with the action of 11--dimensional supergravity~\eqref{sugra}
including the Green--Schwarz term~\eqref{X8}. We would like to
consider this theory on a space--time background of the structure
\begin{equation}
 M_{11}=X\times S^1\times M_4
\end{equation}
where $X$ is a Calabi--Yau three--fold and $M_4$ is four--dimensional
Minkowski space. In addition to the metric background, we would also like
to allow a non--zero mode background for the antisymmetric tensor
field strength $G$ in the internal Calabi--Yau space. Given that $G$
has to be an element of $H^4(X)$ the most general form of such a
background is given by
\begin{equation}
 \bar{G} = 4\p^2\l\,\b_i\n^i\, \e (y)\; .\label{nonzero1}
\end{equation}
Recall here that $\{\n^i\}_{i=1,\dots ,h^{1,1}}$ is a basis of
$H^4(X)$, the constant $\l$ was defined in Eq.~\eqref{lambda} and
$\e (y)$ is the step-function. The coordinate of the circle $S^1$ is called $y$
and we use the same conventions as previously for the orbifold coordinate.

As we have mentioned previously in section \ref{sec:d5het}, the appearance of the
step function $\e (y)$ in the background \eqref{nonzero1} is motivated by the wish
to leave the $y\rightarrow -y$ $Z_2$ symmetry unbroken. Making a generalized
Kaluza-Klein ansatz such as \eqref{nonzero1} without the $\e (y)$ would break the
$Z_2$ symmetry because $G$ is $Z_2$ odd. This choice is also motivated by the wish
to have interesting solutions to the resulting $D=5$ field equations, because these
field equations acquire a cosmological potential term~\cite{losw1,losw2} which
rules out flat space, or indeed any maximally symmetric space, as a solution.
Including the $\e (y)$ in the ansatz \eqref{nonzero1} allows solutions that have
the character of 3-brane domain walls from a $D=5$ perspective, although from a
$D=11$ perspective they appear as 5-branes wrapped around two-cycles of the
Calabi--Yau space and piled into stacks for reduction in the other four
Calabi--Yau directions~\cite{losw1,losw2}. This purely field-theoretic background
solution accords precisely with the Hor\v{r}ava--Witten orbifold structure that
appears to be a non-intrinsic injection into the theory from the
$D=11/D=10$ perspective of section \ref{sec:d5het}. Thus, in $D=5$, the orbifold
becomes a natural aspect of the field-theoretic background.

The form \eqref{nonzero1} for the non--zero mode is identical to the one we 
used in section \ref{sec:d5het}, Eq.~\eqref{nonzero}. An important difference,
however, is that the constants $\b_i$, although quantized as usual, are here taken
to be otherwise undetermined. In section \ref{sec:d5het}, the non--zero mode
was forced upon us due to gauge field and gravity sources in the non--trivial
Bianchi identity of Ho\v rava--Witten construction. Correspondingly, the $\b_i$
were determined in terms of those sources by Eq.~\eqref{bi}. Here,  the non--zero
mode and the coefficients $\b_i$ in particular may be freely chosen. For example,
they could be set to zero if desired. 

The form \eqref{nonzero1} of the Kaluza-Klein ansatz corresponds to having just
two regions in the $y$ coordinate, with equal and opposite non-zero mode charges.
This choice can be extended by the inclusion of additional extended objects
at various $y$ values, by the inclusion of additional step functions in
ans\"atze such as \eqref{nonzero1}. All such cases will be constrained, however,
by a cohomological condition~\cite{nse}
\begin{equation}
\sum_{{\cal S}^1\,{\rm patches}\; (n)}\beta_i^{(n)} = 0\; .\label{cohomcond}
\end{equation}
This is derived by requiring that, although the Bianchi identity for $G$ is
modified by the inclusion of delta--function terms such as in \eqref{bianchi},
the form $dG$ should still be exact in the full $D=11$ spacetime, hence requiring
that its integral vanish when integrated over any closed cycle. The ansatz
\eqref{nonzero1} gives the simplest nontrivial solution to this cohomological
requirement.

The five--dimensional effective theory on $M_5=S^1\times M_4$ obtained by reducing
on the Calabi--Yau space with the generalized ansatz \eqref{nonzero1} is given by
gauged $\cN =1$ supergravity. We are interested here in the parts of this action
involving the antisymmetric tensor field. These are identical to the bulk parts of
the action~\eqref{S5} used in section \ref{sec:d5het} which we repeat here for
convenience. The kinetic and topological terms are given by
\begin{equation}
 2\k_5^2\,S_{\rm kin}=-\int_{M_5}\left[2G_{ij}\cD^i\w
 *\cD^j+2V^{-1}X\w *X^*+V^2G\w *G\right]\label{Skin1}
\end{equation}
and
\begin{multline}
 2\k_5^2\,S_{\rm top} =
  -\sqrt{2}\int_{M_5}\left[\frac{\p^2\l^2}{6v^{2/3}}\;\g_i\,\cB^i\w\tr R^2
  -\frac{8\p^2\l}{v^{2/3}}\e (y)\;\b_i\,\cB^i\w G +\frac{1}{3}d_{ijk}
  \cB^i\cD^j\cD^k\right.\\
  \left. +i(\x G\w X^*-\x^*G\w X)\right]\; .
  \label{Stop1}
\end{multline}
We note that the constant $\g_i$ are proportional to the second Chern
classes of the Calabi--Yau tangent bundle and have been defined in
Eq.~\eqref{gi}. The constants $d_{ijk}$ are the Calabi--Yau intersection
numbers~\eqref{d}. Hence, both $\g_i$ and $d_{ijk}$ are still given to us
by topological properties of the compactification. The non-zero mode
charges $\b_i$ are topological quantities related to the quantized flux of
antisymmetric tensor fields on the internal manifold, but are not related
to the topology of the Calabi-Yau space itself. Another difference with
respect to the previous section concerns the viewpoint one takes with respect to
the modified Bianchi identities. Here, this modification is
seen to follow from the ansatz \eqref{nonzero1}, corresponding to the inclusion of
magnetically charged objects located at the step--function jump points $y=0$,
$y=\pm\p\r$. In section \ref{sec:d5het}, the modifications to the Bianchi identity
were prescribed externally as a consequence of the Ho\v{r}ava--Witten $D=11/D=10$ 
orbifold construction. Note that the modification to the Bianchi identity that we
have so far, just from the background ansatz \eqref{nonzero1}, is still quite
modest, corresponding just to the usual source for a static magnetically charged
extended object without further worldvolume excitations. Moreover, the Bianchi
identity modifications implied by \eqref{nonzero1} have non--vanishing projections
only with four indices in the Calabi-Yau directions, and the remaining one in the
orbifold $y$ direction. The Bianchi identity modifications presented in section
\ref{sec:d5het}, on the other hand, are more extensive, containing also sources
for orbifold-plane modes that we have not yet seen from the $D=5$ field-theoretic
viewpoint, and which have projections with all indices in the $D=5$ space. To see
the r\^ole of such modes, we now turn to a closer inspection of the modified
Bianchi identities.

\subsection{Orbifolding and modifying the Bianchi identities}

We would now like to consider solutions to the above $D=5$ theory having the
space--time character
\begin{equation}
 M_5=S^1/Z_2\times M_4
\end{equation}
that is including a circle orbifolded by $Z_2$. Our conventions here are the
same as those described in the beginning of section 2.1. We shall accordingly have
two four--dimensional orbifold planes $M_4^{(n)}$, where $n=1,2$, located
at the step--function jump points $y=y_1\equiv 0$ and $y=y_2\equiv\p\r\sim-\p\r$.
It is then easy to see from~\eqref{Stop1} that $\cB^i$ and $G$ have to be $Z_2$ odd
fields~\footnote{We call an antisymmetric tensor field even (odd) if
its components orthogonal to the orbifold are even (odd).}. We already
know that there exists a static BPS double domain wall solution of the bulk
supergravity with the three--brane domain walls identified with the two
orbifold planes~\cite{losw1,losw2}.

Let us now assume the existence of ``twisted'' states on these orbifold
planes. The five--dimensional bulk gravitino, corresponding to eight
states, is subject to a chirality constraint on the orbifold planes.
Hence, on these planes we have four--dimensional $\cN =1$
supersymmetry corresponding to four supercharges. We start by assuming a
set of $\cN =1$ gauge fields on each plane with associated gauge groups
\begin{equation}
 \cG_n=\cH_n\times \cJ_n\; .
\end{equation}
Here $\cH_n$ is the semi--simple part and the $U(1)$ factors are
collected in $\cJ_n$. We denote $\cH_n$ gauge fields by
$A_n^\a$ with field strength $F_n^\a$ where $\a$ runs over the simple
factors. Furthermore the $U(1)$ fields are denoted by $\cA_n^a$ with
field strengths $\cF_n^a$ where $a$ labels the various
$U(1)$ factors. Also we assume the existence of $\cN =1$ chiral multiplets
on each plane transforming in the representations $L_n^r$ of $\cG_n$.
Of course the configuration we choose cannot be completely arbitrary
since the full theory has to be anomaly--free. More precisely,
anomalies should cancel on each plane individually. While this can be
done, of course, by choosing a field content with vanishing triangle
anomalies on each plane we know from the previous experience that this is
not the most general case. We may allow for some non--vanishing
triangle anomalies provided they can be cancelled by anomaly inflow
from the bulk. We know that such an inflow should be generated by a
non--trivial variation of the topological part~\eqref{Stop1} of the
action due to a gauge transformation of the antisymmetric tensor
fields. Such a gauge transformation may originate from a non--trivial
Bianchi identity with sources supported on the orbifold planes and
depending on orbifold gauge fields. To incorporate this possibility,
we would like to modify the Bianchi identity in the most general way.
We write
\bea
  dG &=& -\l\sum_{n=1}^2J_n\w dy\,\d(y-y_n) \label{b1m}\\
  d\cD^i &=& -2v^{-1/3}\l\sum_{n=1}^2I_n^i\w dy\,\d(y-y_n)
            \label{b3m}
\eea
with four--forms $J_n$ and two--forms $I_n^i$ which reside on the
orbifold plane. They should be given in terms of orbifold gauge fields
eventually but are kept arbitrary for the moment. Note that we have
not considered the Bianchi identity for the field $\x$ here. The
reason is that the smallest gauge--invariant form on the orbifold
planes is a two--form which does not fit on the right hand side of the
$\x$ Bianchi identity. We also write the integrated form of the above
identities
\bea
  G &=& dC-\l\sum_{n=1}^2w_n\w dy\d(y-y_n) \label{bi1m}\\
 \cD^i &=& d\cB^i-2v^{-1/3}\l\sum_{n=1}^2v_n^i\w dy\d(y-y_n)
            \label{bi3m}
\eea
where the ``Chern--Simons forms'' $w_n$ and $v_n^i$ satisfy
\begin{equation}
 dw_n=J_n\; ,\qquad dv_n^i=I_n^i\; .
\end{equation}

\subsection{Variation of the action}

To compute the classical gauge variation of the action, we first note
that the topological part of the action~\eqref{Stop1} depends on $C$
only through its field strength $G$. As in the previous section, we
are therefore only concerned with the transformation of $\cB^i$.
We start with some transformation law
\begin{equation}
 \d v_n^i=d\l_n^i \label{trafo}
\end{equation}
for the forms $v_n^i$, where the $\l_n^i$ are transformation
parameters to be determined later. Requiring $\d\cD^i=0$, we
find from Eq.~\eqref{bi3m} that
\begin{equation}
 \d\cB^i=2v^{-1/3}\l\sum_{n=1}^2\l_n^idy\,\d (y-y_n)\; .
\end{equation}
To compute the variation of the action we will also need the behavior
of the field strengths close to the orbifold planes. From
Eqs.~\eqref{b1m} and \eqref{b3m} and the associated equations of
motion one finds
\bea
 dy\w G\left.\right|_{y=y_n} &=& \mp\frac{\l}{2}J_n\w dy \\
 dy\w\cD^i\left.\right|_{y=y_n} &=& \mp\l v^{-1/3}I_n^i\w dy\; .
\eea
The classical variation of the bulk action is then given by
\begin{equation}
 \d_{\rm cl}S_5 =
 -\frac{c^3}{128\p^3}\sum_{n=1}^2\int_{M_4^{(n)}}\l_n^i
 \left[\frac{1}{12}\g_i\tr R^2\mp 2\b_iJ_n+d_{ijk}I_n^j\w
 I_n^k\right]\; .\label{dS0}
\end{equation}
To proceed further, we obviously need to know more about the
gauge field parts of the sources $J_n$ and $I_n^i$. Its
most general form is given by an arbitrary linear combination of
all gauge invariant forms of the correct degree. This leads
to
\bea
 J_n &=& \sum_\a k_{n\a}\tr (F_n^\a)^2+\sum_{a,b}h_{nab}\cF_n^a\cF_n^b-l_n\tr
 R^2 \label{an1}\\
 I_n^i &=& \eta_{na}^i\cF_n^a\label{an2}
\eea
where $k_{n\a}$, $h_{nab}$, $l_n$ and $\eta_{na}^i$ are arbitrary
constants. Here, the index $\a$ runs over the various simple groups
contained in the semi--simple part $\cH_n$ of the gauge group.
Writing the transformation parameters $\l_n^i$ as
\begin{equation}
 \l_n^i=\eta_{na}^i\L_n^a \label{red}
\end{equation}
with new parameters $\L_n^a$ we learn from Eq.~\eqref{an2} that our
initial transformation~\eqref{trafo} actually describes the
$U(1)$ gauge variation
\begin{equation}
 \d\cA_n^a=d\L_n^a\label{dAk}\; .
\end{equation}
Hence, similarly as in the previous section, the bulk vector
fields $\cB^i$ transform under the $U(1)$ gauge
symmetries only. This also implies that triangle anomalies not
involving those $U(1)$ field must be zero as they cannot be cancelled by
anomaly inflow. In this sense, the semi--simple parts $\cH_n$ of the
gauge groups must be anomaly--free on both orbifold planes. Depending
on the coefficients $\eta_{na}^i$ in Eq.~\eqref{an2} some of the
$U(1)$ gauge fields might not enter the Bianchi identity~\eqref{b3m}.
To bring our conventions in line with the previous section, we will call such
$U(1)$ fields to be of type II and add them to the semi--simple parts $\cH_n$
of the gauge group. All $U(1)$ fields that do contribute to the
Bianchi identity will be called type I and remain in the $\cJ_n$ part.

Then, inserting the above Ans\"atze into the variation~\eqref{dS0} we arrive
at
\begin{multline}
 \d_{\rm cl}S_5 =
 -\frac{c^3}{128\p^3}\sum_{n=1}^2\int_{M_4^{(n)}}\L_n^a
 \left[(\mp\b_i\eta_{na}^i)\sum_\a k_{n\a}\tr (F_n^b)^2
 -\left(\mp 2l_n\b_i-\frac{1}{12}\g_i\right)\eta_{na}^i\tr R^2 \right.\\
 \left. +\left(\mp 2\b_i\eta_{na}^ih^{bc}_n+\frac{1}{6\p^2}d_{ijk}\eta_{na}^i
 \eta_{nb}^j\eta_{nc}^k\right)\cF_n^b\cF_n^c\right]\label{dS1}
\end{multline}
for the classical variation of the action. The meaning of the three
terms above is obvious. They correspond to a mixed $U_I(1)$ gauge
anomaly, a mixed $U_I(1)$ gravitational anomaly and a $U_I(1)$ cubic
anomaly. This expression has to be compared with the quantum
variation due to triangle diagrams on each boundary. Let us define
the triangle anomaly coefficient $\cC$ as in
Eqs.~(\ref{cd1}a--\ref{cd6}f). Unlike in the previous section,
we have no further information about the particle content on the
orbifold plane. We simply require, at this point, the associated
triangle anomaly to be such that it is cancelled by the anomaly inflow
\eqref{dS1}. This gives the following specific form for the anomaly
coefficients
\begin{subeqnarray}
 \cC_n &=& 0 \label{ck1} \\
 \cC_{na,\a} &=& \mp\frac{1}{8\p}\eta_{na}^i\b_ik_{n\a} \label{ck2} \\
 \cC_{nab} &=& 0 \label{ck3} \\
 \cC_{nabc} &=& \frac{3}{8\p}\left[\mp \b_i\eta_{n(a}^ih_{nbc)}+
                \frac{1}{12\p^2}d_{ijk}\eta_{na}^i\eta_{nb}^j
                \eta_{nc}^k\right] \label{ck4} \\
 \cC^{(L)}_n &=& 0 \label{ck5} \\
 \cC^{(L)}_{na} &=& \frac{3}{2\p}\left(\mp2l_n\b_i-\frac{1}{12}\g_i\right)
                    \eta_{na}^i\; .\label{ck6}
\end{subeqnarray}
The associated quantum variation of the gauge theories on the
orbifold planes then takes the form
\begin{equation}
 \d_{\rm Q}S_{\rm bound} =\frac{1}{16\p^2}\sum_{n=1}^2\int_{M_4^{(n)}}
 \L_n^a\left[\sum_\a C_{na,\a}\tr (F_n^\a )^2-\frac{1}{24}C_{na}^{(L)}\tr
 R^2 +\frac{1}{3}C_{nabc}\cF_n^b\cF_n^c\right]\; .\label{dSQm}
\end{equation}

\subsection{Discussion}

Let us now discuss our results. We would like to compare the general
structure of anomalies, as given in Eqs.~(\ref{cd1}a--\ref{cd6}f), with
the specific form (\ref{ck1}a--\ref{dSQm}) that we have found by
considering five--dimensional gauged supergravity (as obtained from
11--dimensional supergravity) on an orbifold. Also, we would like to
compare those constraints to the corresponding
ones (\ref{c1}a--\ref{c6}f) that we have obtained within heterotic
M--theory. As far as the relation of the two orbifold planes is
concerned, the general discussion in subsection 3.3 applies. Anomaly
cancellation works independently on both planes and the subsequent
discussion should be applied to either one of them.

Let us first discuss the vanishing anomaly coefficients in our list. From
Eq.~(\ref{ck1}a--\ref{ck6}f) those are the cubic anomalies for the
$\cH_n$ part of the gauge group, the mixed gravitational anomaly of $\cH_n$
and the mixed anomaly of one $\cH_n$ field with two
$U(1)$ fields. In particular, we conclude that the semi--simple part of the
orbifold gauge groups always has to be anomaly--free.  This structure of
vanishing coefficients is the same as in heterotic M--theory as given in
Eqs.~(\ref{c1}a--\ref{c6}f). In both cases we remain with three types of
non--vanishing coefficient. While the structure of those remaining
coefficients is similar, there are also some significant differences. To
discuss these it is useful to distinguish various classes of parameters
that determine those coefficients. First, there is topological data of the
Calabi--Yau space, namely the second Chern class of the Calabi--Yau tangent
bundle $\g_i$ and the intersection numbers $d_{ijk}$. Clearly this data is
determined in terms of the underlying compactification in both cases and it
enters the respective formulae in a very similar way. Secondly, there are
the coefficients $\b_i$ and $\eta_{na}^i$. They appear in a very similar way
in the anomaly coefficients, but their interpretation is different in the
two cases. While they are determined in terms of topological data of
the internal vector bundles
in the case of heterotic M--theory, they are merely just parameters here.
While the $\b_i$ parameterize the non--zero mode that we have put in to
compactify 11--dimensional supergravity, the $\eta_{na}^i$ appear as free
parameters in our Ansatz~\eqref{an2} for the sources in the Bianchi
identities. Thirdly, we have the parameters $k_{n\a}$, $h_{nab}$ and $l_n$
that appear in the Ansatz~\eqref{an1} for the sources. Those parameters
arise because of our lack of knowledge about the precise form of the
Bianchi identity for the four--form $G$ in the present case. In heterotic
M--theory they were determined exactly, as comparison of the
Ansatz~\eqref{an1} with Eq.~\eqref{Jn} shows. 

\vspace{0.4cm}

As a result, we have a number of significant differences with respect to
the earlier discussions. Some ``numerical freedom'' in the anomaly
coefficient as compared to heterotic M--theory is introduced by the
arbitrariness of the second group of parameters, that is, $\b_i$ and
$\eta_{na}^i$. However, given that one cannot completely classify the
allowed values for those parameters even for heterotic M--theory (as they
depend, for example, on the choice for the Calabi--Yau manifold) it is hard
to quantify this difference. More significant are some of the differences
related to parameters in the third class. We note, however, that $h_{nab}$
can be absorbed into a redefinition of the $U(1)$ fields $\cA_n^a$. While
the parameters $l_n$ make the mixed gravitational anomaly (\ref{ck6}f) more
flexible, the crucial difference comes from the parameters $k_{n\a}$ in
Eq.~(\ref{ck2}b). Recall, here, that the index $\a$ labels the various
simple groups in $\cH_n$ and, according to our above convention, also the
various $U_{II}(1)$ factors of type II in $\cH_n$. The presence of the
parameters $k_{n\a}$ indicates that the mixed $U_I(1)$ gauge
anomaly (\ref{ck2}b) can be different for each of those factors. Such a
non--universal mixed anomaly is rather different from what we found in
heterotic M--theory (compare with Eq.~(\ref{c2}b) where it was always
gauge--factor independent.  Finally, a crucial difference concerns the size
of the orbifold gauge groups. While, within the context of heterotic
M--theory, they are bound to fit within $E_8$ on each plane no such
restriction arises in the present context. 

\vspace{0.4cm}

To summarize, we have found anomaly cancellation in the setting of this
section to be more flexible than in five--dimensional heterotic M--theory
in two important ways. Firstly, non--universal mixed $U(1)$ gauge anomalies
are possible. Secondly, the orbifold groups are not restricted to fit into
$E_8$. It is clear that both generalizations can be of considerable
phenomenological importance. An important question is whether such
generalizations have an interpretation in terms of M--theory. It would be
very interesting to search for such an interpretation, for example in the
context of heterotic M--theory models with
five--branes~\cite{nse,dlow,dlow1,lows}. However, we will not pursue this
further in the present paper.

%%%%%%%%%%%%%%%%%%%%%%%%%%%%%%%%%%%%%%%%%%%%%%%%%%%%%%%%%%%%%%%%%%%%%%%%%%%%%

\section{Heterotic anomaly cancellation in four dimensions}

%%%%%%%%%%%%%%%%%%%%%%%%%%%%%%%%%%%%%%%%%%%%%%%%%%%%%%%%%%%%%%%%%%%%%%%%%%%%%

In section 3 we have analyzed $E_8\times E_8$ heterotic anomaly
cancellation from a five--dimensional viewpoint. However, upon dimensional
reduction five--dimensional heterotic M--theory reproduces the effective
four--dimensional action of the $E_8\times E_8$ heterotic
string~\cite{losw2}, at least to one--loop order. Hence, the results for
the structure of anomaly cancellation that we found in five dimensions
should be directly applicable to the four--dimensional theory. This seems
rather puzzling, however, since this structure contradicts in various ways
what is commonly assumed about anomaly cancellation in the
four--dimensional $E_8\times E_8$ heterotic string. Usually, it is stated
that there is at most one anomalous $U(1)$ symmetry that ``extends'' over
the hidden and the observable sector. Its quantum anomaly is supposed to be
cancelled exclusively due to a non--trivial transformation law of the
dilaton superfield. As a consequence of the universal dilaton coupling, in
order for such a cancellation to work, all triangle anomaly coefficients
(including the cubic and the mixed gravitational one) have to be in fixed
proportions. On the other hand, the structure that we found in five
dimensions turned out to be significantly more flexible. Various anomalous
$U(1)$ symmetries were possible, hidden and observable sectors were greatly
independent, and  some of the triangle anomaly coefficients were
generically unrelated. Moreover, the cancellation mechanism was due to a
transformation of the five--dimensional vector fields, which, from a
four--dimensional viewpoint, are associated with the $T$ moduli rather than
with the dilaton. As we will see, these discrepancies are resolved by a
careful distinction between $U(1)$ symmetries of type I and type II and
their properties. Firstly, however, we would like to study
four--dimensional anomaly cancellation systematically in order to confirm
our previous results and gain some confidence. The effective
four--dimensional action that we are going to need for this can be obtained
in two ways. Firstly one can reduce the 10--dimensional $E_8\times E_8$
heterotic action on a Calabi--Yau three--fold to four dimensions. Secondly,
we can start with the five--dimensional heterotic M--theory as given in
section 2 and reduce it to four dimensions. Both methods have to agree
since the five-- and the four--dimensional theories, as well as
the 11-- and 10-- dimensional theories are equivalent. This
has been explicitly verified in references~\cite{losw2} and \cite{elten},
respectively. Here we will use the first approach, if only to have a better
comparison to the weakly coupled heterotic string.

\subsection{The 10--dimensional action}

We are interested in the part of the 10--dimensional $E_8\times E_8$
heterotic effective action that involve the two--index antisymmetric tensor
field $B$ with field strength $H=dB+\cdots$ as well as certain terms involving
the curvature and the $E_8\times E_8$ gauge fields $A$ with fields
strength $F$. This part of the action is given by
\begin{equation}
 S_{10} = S_{\rm kin}+S_{\rm top}
\end{equation}
where
\bea
 S_{\rm kin} &=& -\frac{1}{2\k_{10}^2}\int_{M_{10}}\left[
                 e^{-2\f}H\w *H+\frac{\a '}{4}(\tr F^2-\tr
                 R^2)\right]\\
 S_{\rm top} &=& -k\int_{M_{10}}B\w W_8\; .
\eea
Here $\f$ is the dilaton and the constant $k$ is given by
\begin{equation}
 k = \frac{c^3}{3\sqrt{2}\cdot 2^7\p^5\a '} \; .
\end{equation}
The non--trivial Bianchi identity for $H$ reads
\begin{equation}
 dH = -\frac{\a '}{2\sqrt{2}}(\tr F^2-\tr R^2)\; .
\end{equation}
Furthermore, the anomaly polynomial $W_8$ has the well--known form
\begin{equation}
 W_8 = \frac{1}{24}\Tr F^4-\frac{1}{7200}\left(\Tr F^2\right)^2
       -\frac{1}{240}\Tr F^2\tr R^2+\frac{1}{8}\tr R^4+
       \frac{1}{32}\left(\tr R^2\right)^2\; .\label{W8}
\end{equation}
As usual, $\Tr$ denotes the trace in the adjoint, while $\tr = \Tr
/30$. It is useful for our purpose to express the anomaly polynomial in
terms of the individual $E_8$ gauge fields $A_n$ with field strengths
$F_n$. Writing $F=F_1+F_2$, one obtains
\begin{equation}
 W_8 =\frac{1}{4}\left(\tr F_1^2\right)^2+\frac{1}{4}
      \left(\tr F_2^2\right)^2-\frac{1}{4}\tr F_1^2\tr F_2^2
      -\frac{1}{8}\left(\tr F_1^2+\tr F_2^2\right)\tr R^2
      +\frac{1}{8}\tr R^4+\frac{1}{32}\left(\tr R^2\right)^2\; .
      \label{W81}
\end{equation}

\vspace{0.4cm}

We would like to reduce the above theory on a background 
space--time
\begin{equation}
 M_{10}=X\times M_4
\end{equation}
where $X$ is a Calabi--Yau three--fold with metric\footnote{Here and
in the following, the index $10$ refers to quantities measured in the
10--dimensional string metric, as opposed to the 11--dimensional
Einstein metric used earlier.} $\bar{g}_{10}$ and
curvature two--form $\bar{R}$. As far as the gauge fields are
concerned, the background configuration is exactly as described for the
reduction of Ho\v rava--Witten theory. We will not repeat this here
but simply refer back to section 2.2. 

Let us list the relevant zero modes around this background. We define
the breathing modulus $V$ to be
\begin{equation}
 V = \frac{e^{-2\f}}{v}\int_X\sqrt{\bar{g}_{10}}\; .
\end{equation}
The $(1,1)$ moduli $a^i_{10}$ appear in the expansion of the K\"ahler
form $\O_{10}=a^i_{10}\O_i$ in terms of the basis
$\{\O_i\}_{i=1,\cdots ,h^{1,1}}$ of harmonic $(1,1)$ forms. For the
antisymmetric tensor field and its field strength we write
\bea
 B &=& \tilde{B}+\c^i\O_i \\
 H &=& \tilde{H}+X^i\w\O_i
\eea
where $\tilde{B}$ is the four--dimensional two--form with field
strength $\tilde{H}$ and $\c^i$, where $i=1,\cdots ,h^{1,1}$ are
axionic scalars with field strengths $X^i$. We have dropped
contributions from $(2,1)$ modes which are not important in our context.
In terms of four--dimensional multiplets, $V$ and the dual of
$\tilde{B}$ form the bosonic part of the dilaton superfield $S$,
whereas $a_{10}^i$ and $\c^i$ represent the bosonic field content of
the $T^i$ moduli. For the gauge field zero modes originating from
$E_8\times E_8$, exactly the same discussion as in the 11--dimensional
case applies. We adopt the notation of section 2.3 to which we refer
back for details.

\subsection{The four--dimensional effective action}

By straightforward calculation, using the above setup, we find 
the part of the four--dimensional effective action relevant
for anomaly cancellation
\begin{equation}
 S_4 = S_{4,{\rm kin}}+S_{4,{\rm top}}\label{S4}
\end{equation}
where the kinetic part is given by
\begin{multline}
 S_{4,{\rm kin}} = -\frac{1}{2\k_4^2}\int_{M_4}\left[V^2H\w
    *H+2G_{ij}X^i\w *X^j\right] \\
    -\frac{1}{4g_0^2}\int_{M_4}\sqrt{-g}
    V\left[\tr F_1^2+\tr F_2^2+\sum_A(\cF^A)^2\right]\; .\label{S4kin}
\end{multline}
The topological part reads
\begin{equation}
 S_{4,{\rm top}} = -6\p^2 k\int_{M_4}\left[ 2\b_i\hat{\eta}_A^i B\w\cF^A
                   +\c^i\left( -\b_i\tr F_1^2+\b_i\tr F_2^2
                   +f_{iAB}\cF^A\cF^B+\frac{1}{12}\g_i\tr
                   R^2\right)\right]\; .\label{S4top}
\end{equation}
 From the reduction of the 10--dimensional Bianchi identity we have
for the field strengths
\bea
 H &=& dB-\frac{\a '}{2\sqrt{2}}\o_3 \label{biH}\\
 X^i &=& d\c^i-\frac{\a '}{\sqrt{2}v^{1/3}}\eta_A^i\cA^A \label{biX}
\eea
with the Chern--Simons form $\o_3$ satisfying
\begin{equation}
 d\o_3 = \tr F_1^2+\tr F_2^2+\sum_A(\cF^A)^2-\tr R^2\; .\label{w3}
\end{equation}
To simplify the notation, we have used indices $A,B,\cdots$ to denote
the two $E_8$ factors as well as the various $U_I(1)$ factors.
So, for example, $A=(na)$ where $n=1,2$ and $a$ runs over the
$U_I(1)$ factors in each sector. Correspondingly, we have defined the
following short--hand notation 
\begin{equation}
 \eta_A^i = (\eta_{1a}^i,\eta_{2a}^i) \; ,\qquad
 \hat{\eta}_A^i = (-\eta_{1a}^i,\eta_{2a}^i)
 \label{etaA}
\end{equation}
and
\begin{equation}
 f_{iAB} = \left(\begin{array}{cc}
           \frac{1}{6\p^2}d_{ijk}\eta_{1a}^j\eta_{1b}^k-\b_i\d_{ab}&
           -\frac{1}{12\p^2}d_{ijk}\eta_{1a}^j\eta_{2b}^k \\
           -\frac{1}{12\p^2}d_{ijk}\eta_{1a}^j\eta_{2b}^k&
          \frac{1}{6\p^2} d_{ijk}\eta_{2a}^j\eta_{2b}^k+\b_i\d_{ab}
           \end{array}\right)\label{fdef}
\end{equation}
for the various coupling constants in the effective action. 
The four--dimensional Newton constant $\k_4$ and the gauge coupling
$g_0$ are defined as
\begin{equation}
 \k_4^2 = \frac{\k_{10}^2}{v}\; ,\qquad g_0^2 = \frac{2\k_{10}^2}{v\a
 '}\; .
\end{equation}
It is useful to recall the structure of the gauge fields. The
low--energy gauge group $\cG_n$ in each sector is given by the product
\begin{equation}
 \cG_n=\cH_n\times\cJ_n\; ,
\end{equation}
where $\cH_n$ contains the semi--simple part and the $U_{II}(1)$ generators
(which are not part of the internal structure
group) while $\cJ_n$ contains the $U_I(1)$ generators (which
are part of the internal structure group).
Gauge fields in $\cH_n$ are denoted by $A_n$ with field strengths
$F_n$. The $U_I(1)$ fields of type I are denoted by $\cA_n^a$ with field
strengths $\cF_n^a$ or $\cA^A$ and $\cF^A$ in the above index convention.
We also recall that the topological numbers $\b_i$ and $\g_i$, related
to the second Chern characters of the internal bundles, have been
defined in Eq.~\eqref{bi} and \eqref{gi}, respectively. Furthermore,
$d_{ijk}$ are the Calabi--Yau intersection numbers~\eqref{d}.

\subsection{Classical variation of the action}

We would now like to compute the classical gauge variation of the
above action. We introduce the transformation parameters $\L^A$ for
the $U_I(1)$ fields $\cA^A$ and the transformation parameters
$\L_n$ for the gauge fields $A_n$ in $\cH_n$. Furthermore, we will
need the spin connection $w$ and the associated transformation
parameter $\L_L$. We write explicitly
\bea
 \d\cA^A &=& d\L_A \\
 \d A_n &=& d\L_n+A_n\L_n-\L_nA_n \\
 \d w &=& d\L_L +w\L_L-\L_Lw\; .
\eea
It is useful to note the resulting transformation of the Chern--Simons
form $w_3$ defined in Eq.~\eqref{w3}. It is given by
\begin{equation}
 \d w_3 = \sum_A\L^A\cF^A+\sum_{n=1}^2d\,\tr (\L_ndA_n)-d\,\tr (\L_Ldw)\; .
\end{equation}
As before, the non--trivial gauge transformations of antisymmetric
tensor fields are triggered by the Bianchi identities~\eqref{biH} and
\eqref{biX}. Requiring the field strengths $H$ and $X^i$ to be gauge
invariant, we find the transformation laws
\bea
 \d B &=& \frac{\a '}{2\sqrt{2}}\left(\sum_A\L^A\cF^A+
          \sum_{n=1}^2\tr (\L_ndA_n)-\tr (\L_Ldw)\right) \label{dB}\\
 \d\c^i &=& \frac{\a '}{\sqrt{2}}\eta_A^i\L^A\; .\label{dc}
\eea
This leads to the following classical variation of the action
\begin{multline}
 \d_{\rm cl}S_4 = -3\sqrt{2}\p^2k\a '\int_{M_4}\bigg\{\L^A
                    \bigg[\eta_A^i\left( -\b_i\tr F_1^2+\b_i\tr F_2^2
                    +\frac{1}{12}\g_i\tr R^2\right) \\
                    +\left(\eta_A^if_{iBC}+\b_i\hat{\eta}_C^i
                    \d_{AB}\right)\cF^B\cF^C\bigg] \\
                +\b_i\hat{\eta}_A^i\left[\tr (\L_1dA_1)+
                 \tr (\L_2dA_2)-\tr (\L_Ldw)\right]\cF^A\bigg\}\; .
\end{multline}
To understand the structure of this variation and to compare it to
the previous results we should write it in a more explicit form and split
all terms explicitly into the two sectors. Doing this, one arrives at the
somewhat uncomfortable form
\begin{subeqnarray}
 \d_{\rm cl}S_4 &=&-3\sqrt{2}\p^2k\a '\int_{M_4}\bigg\{\nn \label{dS4}\\
                 &&\L_1^a\bigg[ -\b_i\eta_{1a}^i\tr F_1^2
                   +\b_i\eta_{1a}^i\tr F_2^2+\frac{1}{12}
                   \g_i\eta_{1a}^i\tr R^2 \label{t1} \\
                 &&\qquad +\left( -\b_i\eta_{1a}^i\d_{bc}-\b_i\eta_{1c}^i
                   \d_{ab}+\frac{1}{6\p^2}d_{ijk}\eta_{1a}^i\eta_{1b}^j
                   \eta_{1c}^k\right)\cF_1^b\cF_1^c \label{c11} \\
                 &&\qquad+\left(\b_i\eta_{2c}^i\d_{ab}-\frac{1}{6\p^2}d_{ijk}
                   \eta_{1a}^i\eta_{1b}^j\eta_{2c^k}\right)\cF_1^b\cF_2^c\\
                 &&\qquad +\left(\b_i\eta_{1a}^i\d_{bc}+\frac{1}{6\p^2}d_{ijk}
                   \eta_{1a}^i\eta_{2b}^j\eta_{2c^k}\right)\cF_2^b\cF_2^c
                   \bigg]\label{c13}\\
                 &&\L_2^a\bigg[ -\b_i\eta_{2a}^i\tr F_1^2
                   +\b_i\eta_{2a}^i\tr F_2^2+\frac{1}{12}
                   \g_i\eta_{2a}^i\tr R^2\label{t2} \\
                 &&\qquad +\left( +\b_i\eta_{2a}^i\d_{bc}-\b_i\eta_{2c}^i
                   \d_{ab}+\frac{1}{6\p^2}d_{ijk}\eta_{2a}^i\eta_{2b}^j
                   \eta_{2c}^k\right)\cF_2^b\cF_2^c \label{c21} \\
                 &&\qquad+\left( -\b_i\eta_{2b}^i\d_{ac}-\frac{1}{6\p^2}d_{ijk}
                   \eta_{2a}^i\eta_{1b}^j\eta_{2c^k}\right)\cF_1^b\cF_2^c\\
                 &&\qquad +\left( -\b_i\eta_{2a}^i\d_{bc}+\frac{1}{6\p^2}
                   d_{ijk}\eta_{2a}^i\eta_{1b}^j\eta_{1c^k}\right)
                   \cF_1^b\cF_1^c\bigg]\label{c23}\\
                 &&+\tr (\L_1dA_1)\bigg[ -\b_i\eta_{1a}^i\cF_1^a+
                   \eta_{2a}^i\b_i\cF_2^a\bigg] \label{t3} \\
                 &&+\tr (\L_2dA_2)\bigg[ -\b_i\eta_{1a}^i\cF_1^a+
                   \eta_{2a}^i\b_i\cF_2^a\bigg] \label{t4} \\
                 &&+\tr (\L_Ldw)\bigg[ \b_i\eta_{1a}^i\cF_1^a-
                   \eta_{2a}^i\b_i\cF_2^a\bigg]\bigg\} \label{t5}
\end{subeqnarray}
for the classical variation.

\subsection{Quantum variation of the action}

The four--dimensional effective action that we have computed should be
anomaly--free in the same way as the 10--dimensional action that it
originates from is. As a consequence, the classical variation of the
action computed above should cancel the quantum variation due to
triangle diagrams. As a check we would like to verify this explicitly.

The triangle anomaly coefficients have already been generally defined in
Eqs.~(\ref{cd1}a--\ref{cd6}f). In fact, their specific form has to be
exactly the same as for the five--dimensional heterotic M--theory, since they
depend on the particle content in the gauge sectors only. We can,
therefore, just use the result from section 3.2 which we repeat here
for convenience
\bea
 \cC_n &=& 0 \label{cr1} \\
 \cC_{na} &=& \mp\frac{1}{8\p}\eta_{na}^i\b_i \label{cr2} \\
 \cC_{nab} &=& 0 \label{cr3} \\
 \cC_{nabc} &=& \frac{3}{8\p}\left[\mp \b_i\eta_{n(a}^i\d_{bc)}+
                \frac{1}{12\p^2}d_{ijk}\eta_{na}^i\eta_{nb}^j
                \eta_{nc}^k\right] \label{cr4} \\
 \cC^{(L)}_n &=& 0 \label{cr5} \\
 \cC^{(L)}_{na} &=& \frac{3}{2\p}\left(\mp\b_i-\frac{1}{12}\g_i\right)
                    \eta_{na}^i\; .\label{cr6}
\eea
The upper (lower) sign refers to the $n=1$ ($n=2$) sector. In five
dimensions, the quantum variation was split into two parts, one on each
orbifold plane. The four--dimensional quantum variation is, of course,
just the sum of these two parts. From Eq.~\eqref{dQS5} it is given by
\begin{equation}
 \d_{\rm Q}S_4 = \frac{1}{16\p^2}\sum_{n=1}^2\int_{M_4}\L_n^a\left[
                 \cC_{na}\tr F_n^2-\frac{1}{24}\cC_{na}^{(L)}\tr R^2
                 +\frac{1}{3}\cC_{nabc}\cF_n^b\cF_n^c\right]\label{dQS4}\; .
\end{equation}
Given that this quantum variation was cancelled by the classical variation of the
five--dimensional action~\eqref{dS5}, how can it be cancelled by the significantly
more complicated four--dimensional classical variation~\eqref{dS4}? Again the answer
is related to the ambiguity in the triangle anomaly. The classical
variation~\eqref{dS4} ``wants'' to cancel a quantum variation in a specific
regularization. Comparing the four--dimensional classical variation~\eqref{dS4} with
its five--dimensional counterpart~\eqref{dS5} there are two essential complications.
First, in addition to the anomalous variations of the $U_I(1)$ fields that we
had in five dimensions, the four--dimensional expression contains anomalous variations
of the $\cH_n$ parts of the gauge symmetries and gravity as well. Secondly, whereas in
five dimensions the terms in the anomalous variation were naturally split between the
two orbifold planes, the four--dimensional variation contains terms that mix the two
sectors. 

The first complication can be accounted for by regularizing the mixed
$U_I(1)$ gauge and gravitational anomalies in a non--standard way. Usually, this is done
such that the $U_I(1)$ is anomalous while the rest of the gauge group and gravity are
anomaly free. Here we should adopt a mixed scheme in which all parts become anomalous.
Rewriting the first two terms in the quantum variation~\eqref{dQS4} in such a way,
they can be cancelled by the terms in~(\ref{t1}a), (\ref{t2}e) and
(\ref{t3}i--\ref{t5}k) in the classical variation~\eqref{dS4}. To check that this
works, one has to verify that the coefficients of the corresponding terms in
(\ref{t1}a), (\ref{t2}e) and (\ref{t3}i--\ref{t5}k) add up to what is
required from the standard form of the anomaly~\eqref{dQS4}. This is
indeed the case. What about the cubic
anomaly terms? Here we can use the same ambiguity and rewrite the last term in the
quantum variation to be cancelled by the remaining terms in~\eqref{dS4}. In
particular, in this way one can account for the terms in \eqref{dS4} that mix the two
sectors. To see this, one should add up the coefficients of corresponding terms in
(\ref{c11}b--\ref{c13}d) and (\ref{c21}f--\ref{c23}h). Doing this, we see that
indeed all mixed coefficients cancel. This is essential, because a ``real'' anomaly
that mixes the two sectors would be inconsistent. The two sectors are completely hidden
with respect to one another so there are simply no fermions available that could
generate a mixed sector anomaly. We finally conclude that four--dimensional classical
and quantum variations indeed cancel as they should.

\subsection{Discussion}

What is the structure of four--dimensional anomaly cancellation from
these results? All the essential information can be extracted by
comparing the general definition of the anomaly
coefficients (\ref{cd1}a--\ref{cd6}f) and their special
form~\eqref{cr1}--\eqref{cr6} along with the anomalous (quantum)
variation which we can take to be in its standard form~\eqref{dQS4}.
It is clear, that this leads to exactly the same conclusions as in the
case of heterotic M--theory in five dimensions.

Let us summarize the results from section 3.3 briefly. Recall that
$\cH_n$ consist of the non--abelian and the $U_{II}(1)$ gauge fields,
collectively called $A_n$. Furthermore, we denote the $U_I(1)$ gauge fields by
$\cA_n^a$, where $a=1,\dots ,k_n$ and the graviton is denoted by $g$. We then
refer to a triangle diagram by specifying the triple of gauge fields
to which it couples. Then the anomaly structure in {\em each} sector
is as follows~:
\begin{itemize}
 \item The $\cH_n$ part of the gauge group is anomaly--free in the
        sense that the $A_n^3$ and $A_ngg$ anomalies vanish.
 \item The mixed $\cA_n^a\cA_n^bA_n$ anomalies vanish.
 \item After a suitable choice of basis there is at most a single $U_I(1)$
       gauge field, say $\cA_n^1$ with all three remaining types of
       anomalies non--vanishing. That is, we may have anomalies of
       type $\cA_n^1A_nA_n$, $\cA_n^1\cA_n^a\cA_n^b$ and $\cA_n^1gg$. In terms
       of the topological data, the associated anomaly coefficients are
       given in Eq.~(\ref{c2}b), (\ref{c4}d) and (\ref{c6}f).
       The remaining $U_I(1)$ factors are free of mixed gauge
       anomalies, that is the $\cA_n^aA_nA_n$ anomaly vanishes for $a>1$. 
 \item The coefficient of $\cA_n^1A_nA_n$ is independent of which
       specific gauge field $A_n$ within $\cH_n$ is considered. Apart
       from this restriction, the three non--vanishing anomaly coefficients are
       generically unrelated.
 \item After another choice of basis there is at most one among the
       remaining $U_I(1)$ symmetries, say $\cA_n^2$, with
       mixed gravitational and cubic anomaly. In other words, the anomalies
       of type $\cA_n^2gg$ and $\cA_n^2\cA_n^a\cA_n^b$ can be
       non--vanishing. Again the two associated anomaly coefficients
       are unrelated generically. All other $\cA_n^a$ for $a>2$ have cubic
       anomalies at most.
\end{itemize}

We have seen that the four-- and five--dimensional pictures are
completely consistent with one another. However, the essential
structure is much easier seen in five dimensions where the two sectors
of the theory are located on the orbifold planes and, hence, are ``neatly''
separated. What causes the complication in four dimensions? A
reduction on the orbifold is not a simple truncation of all
fields. In Kaluza-Klein terms, setting all the $y$-dependent modes equal to their
vacuum expectation values (zero or other constant values) does not constitute a
consistent truncation of the $D=5$ theory down to $D=4$.\footnote{The lack of
Kaluza-Klein consistency in the $5\rightarrow 4$ reduction, with consequent impact
upon the $D=4$ interaction structure, should be compared with the ``essentially
consistent'' reduction from $D=11$ to $D=5$ on the Calabi-Yau manifold. In the
latter case, integration of the higher Kaluza-Klein modes of the Calabi-Yau
manifold does not produce corrections in the $D=5$ effective action, at least at
the level of terms containing no more than two derivatives \cite{dfps}.} Instead,
what one has to do is to integrate out all the higher Kaluza-Klein fields of the
bulk theory. This generates ``interactions'' across the orbifold and gives rise,
among other things, to the more complicated structure of anomaly terms. In
particular, the resulting
$D=4$ anomaly structure contains cross terms between the two $E_8$ sectors that
from a purely $D=4$ perspective would not appear to be necessary ({\it i.e.}\ they
could be removed by appropriate regularization or renormalization), but which are
unavoidable here as a result of the higher-dimensional anomaly structure.

\subsection{Dualizing $B$}

Usually, the four--dimensional two--form $B$ is dualized to a scalar
$\s$. While this does not effect the anomalous variation of the
action under gauge symmetries, of course, it is nevertheless instructive to carry this
out explicitly. We write $H_0=dB$ and add the term
\begin{equation}
 \frac{1}{\k_4^2}\int_{M_4}H_0\w d\s
\end{equation}
to the action~\eqref{S4}--\eqref{S4top}. Then, integrating out $H_0$ we find
\begin{multline}
 S_{4,{\rm kin}} = -\frac{1}{2\k_4^2}\int_{M_4}\left[\frac{1}{V^2}\Sigma
    \w *\Sigma +2G_{ij}X^i\w *X^j\right] \\
    -\frac{1}{4g_0^2}\int_{M_4}\sqrt{-g}
    V\left[\tr F_1^2+\tr F_2^2+\sum_A(\cF^A)^2\right]\; .\label{S4kind}
\end{multline}
and
\begin{equation}
 S_{4,{\rm top}} = \frac{\a '}{\sqrt{2}}\k_4^2\int_{M_4}w_3\w\Sigma
                   -6\p^2 k\int_{M_4}\c^i\left( -\b_i\tr F_1^2+
                   \b_i\tr F_2^2+f_{iAB}\cF^A\cF^B+\frac{1}{12}\g_i\tr
                   R^2\right)\; .\label{S4topd}
\end{equation}
where the constants $\eta_A^i$, $\hat{\eta}_A^i$ and $f_{iAB}$ have
been defined in \eqref{etaA}--\eqref{fdef} and the field strengths
$\Sigma$ and $X^i$ are given by
\bea
 \Sigma &=& d\s+12\p^2k\k_4^2\,\b_i\hat{\eta}_A^i\cA^A \label{Sigma}\\
 X^i &=& d\c^i-\frac{\a '}{\sqrt{2}v^{1/3}}\eta_A^i\cA^A\; .\label{Xi1}
\eea
Recall here that $\s$ is the axionic part of the dilaton superfield
$S$ while $\c^i$ are the axions in the $T^i$ moduli. As usual, the two last equations
can be used to deduce gauge transformation laws for the dilaton and the $T$ moduli
(for the $T$ moduli the transformation has been given in Eq.~\eqref{dc}) by requiring
the field strengths to be gauge invariant. We see that also the dilaton transforms
non--trivialy, in general, as did its dual $B$. However, unlike for the $T$ moduli,
the dualized action~\eqref{S4kind}, \eqref{S4topd} is invariant under this
transformation of the dilaton since it depends only on the field strength $\Sigma$. In
particular, the universal coupling of the dilaton to the gauge fields is not given by
$\s dw_3$ but by $\Sigma w_3$. Hence we see that the anomaly cancellation is
essentially controlled by the transformation of the $T$ moduli. This works thanks to
the well--known $T^i$ dependent threshold correction to the gauge kinetic function, of
which the second part of Eq.~\eqref{S4topd} represents the imaginary part.

\subsection{Relation to the usual picture}

Finally, we should discuss how the above results relate to the usual picture of
four--dimensional heterotic anomaly cancellation. To do this, it is useful to review
the standard argument given in the literature~\cite{dsw}. One starts with the
10--dimensional Green--Schwarz term, {\it i.e.}\ roughly with an expression of the form
\begin{equation}
 B\w F^4\; .
\end{equation}
Taking three of the gauge fields to be internal and the remaining one
to be external, this leads to a four--dimensional term of the form
\begin{equation}
 cB\w {\bf F} \label{cc}
\end{equation}
where $B$ is the four--dimensional two--form, ${\bf F}=d{\bf A}$ is a
four--dimensional $U(1)$ gauge field strength and $c$ is some
coefficient. Dualizing $B$ to the axion field $\s$, the above term
becomes
\begin{equation}
 c\partial_\m\s {\bf A}^\m\; .
\end{equation}
In analogy with the previous subsection, this term shows that $\s$
should be transformed under the gauge symmetry associated to ${\bf A}$.
Due to the coupling $\s F^2$ and $\s R^2$ of the dilaton to
all gauge fields and gravitation this then leads to an anomalous
classical variation that should cancel the $U(1)$ triangle anomalies
associated to ${\bf A}$.
 From this line of reasoning a number of properties of the quantum
anomaly can be deduced. It should be universal, that is, it should be
the same for all factors of the gauge group and for gravity. In
particular, this includes the hidden and the observable gauge group.
The reason for this property is the universal coupling of the dilaton.
Furthermore, there is at most one such anomalous $U(1)$ symmetry.
This is the standard reasoning. 

There are two hidden assumptions in the above line of argument. First, it is assumed
that only the dilaton, but not the $T$ moduli, vary under the ${\bf A}$ gauge
symmetry. Secondly, it is assumed that the coefficient $c$ in~\eqref{cc} is non--zero.
The first assumption necessarily implies that the $U(1)$ symmetry in question is of
type II. Otherwise (if it were type I) the $U(1)$ would be part of the internal
structure group. From the $\tr F^2$ terms in the (10--dimensional) Bianchi identity
with one $F$ external and the other internal, this would then lead to a term
proportional to ${\bf A}$ in the $T^i$ Bianchi identities. In other words, ${\bf A}$
would appear on the right hand side of Eq.~\eqref{Xi1} and hence the $T^i$ would have
to transform under the $U(1)$ symmetry. We have assumed that this is not the case and,
hence, we are dealing with a type II symmetry. On the other hand, we have shown that
the type II symmetries are always anomaly free. This must mean that the coefficient
$c$ in Eq.~\eqref{cc} is zero and, therefore, the second assumption above is violated.
In fact, it is easy to understand why $c$ vanishes for type II symmetries~\cite{dg}.
The only term in the anomaly polynomial~\eqref{W8} that could lead to a
coupling~\eqref{cc} is  the first term proportional to $\tr (F^4)$. This is because an
expression of the form $\tr (\bar{F}{\bf F})$ involving the internal field strength
$\bar{F}$ always vanishes if ${\bf F}$ is of type II. On the other hand, it is known
that $E_8$ has no independent quartic invariant. As a consequence, the anomaly
polynomial can be written in terms of second order invariants only, as it has been
done in Eq.~\eqref{W81}. This explains why $c$ is always zero for $U_{II}(1)$
symmetries and, hence, why the coupling~\eqref{cc} does not exist in this
case~\footnote{This is different for the
$SO(32)$ heterotic theory where there exists an independent fourth order invariant. In
fact, the standard embedding leads to an example of an anomalous type II symmetry in
this case~\cite{dsw}.}. To summarize, we have found that one of the assumptions in the
standard argument is not satisfied and, hence, that the conclusions about the
structure of low--energy anomaly cancellation do not apply to the $E_8\times E_8$
case. Essentially, the statement is that the conventional anomaly
properties described above apply to an anomalous type II $U(1)$
symmetry. While for the $SO(32)$ heterotic theory type
I as well as type II $U(1)$ symmetries can be anomalous, we have seen that type II
symmetries are always anomaly--free in the $E_8\times E_8$ case. Using the standard
(type II) assumptions about anomalous $U(1)$ symmetries for $E_8\times E_8$ means,
therefore, applying them to an empty set. On the other hand, we have seen that type I
$U(1)$ symmetries can be anomalous in the $E_8\times E_8$ case and that their
properties are quite different from the usual ones.

%%%%%%%%%%%%%%%%%%%%%%%%%%%%%%%%%%%%%%%%%%%%%%%%%%%%%%%%%%%%%%%%%%%%%%%%%%%%%%

\section{Phenomenological issues}

%%%%%%%%%%%%%%%%%%%%%%%%%%%%%%%%%%%%%%%%%%%%%%%%%%%%%%%%%%%%%%%%%%%%%%%%%%%%%%

\subsection{Model building with anomalous $U(1)$ symmetries}

 From the perspective of low--energy model building it is an important
question as to which anomalous $U(1)$ symmetries (with the quantum anomaly
being cancelled by the Green--Schwarz mechanism) can be added to the
MSSM (or extensions thereof) within the context of heterotic string--
or M--theory effective actions. This question is answered by our
results for the anomaly coefficients. We remind the reader, that, within
each sector, we have split our low--energy gauge groups $\cG_n$,
where $n=1,2$ labels the observable and the hidden sectors, into two
parts, namely
\begin{equation}
 \cG_n =\cH_n\times \cJ_n\; 
\end{equation}
Here $\cH_n$ contains the complete semi--simple part of the gauge
group as well as the $U(1)$ symmetries of type II. The $\cJ_n$ part,
on the other hand, contains the $U(1)$ factors of type I. The
generators of $\cH_n$ are generically called $T_n$, with the associated
gauge fields $A_n$. The $\cJ_n$ generators are denoted by $Q_n^a$ with
gauge fields $\cA_n^a$, where $a$ runs over the various
$U(1)$ factors in $\cJ_n$. In Eqs.~(\ref{cd1}a--\ref{cd6}f) we have
defined in general all possible anomaly coefficients between those
two parts of the gauge group and also with gravity.
Their specific forms within $E_8\times E_8$ models, in terms of data
related to the underlying compactification, has been given
in~\eqref{cr1}--\eqref{cr6}. It is these latter equations that
constitute the crucial input for anomalous $U(1)$ model building in
the ``bottom up'' approach. Recall that, from those coefficients,
the $\cH_n$ parts of the
gauge group are anomaly--free while the $\cJ_n$ contain the
potentially anomalous $U(1)$ symmetries. The generic structure of
the anomalous $U(1)$ symmetries in $\cJ_n$ that results has been
discussed in section 5.5. Here we would like to comment on a few
specific aspects conceivably relevant for model building.

A general feature of our structure is the independence of the
observable and the hidden sector. For example, one may have two
anomalous $U(1)$ symmetries, one in each sector, with generically
unrelated anomaly coefficients. This could conceivably be important for
relevant hidden sector physics such as supersymmetry breaking.
We also gain some flexibility in the observable sector. The anomaly
coefficients for the mixed gauge anomaly ($\cA_n^aA_nA_n$ diagrams), the
mixed gravitational anomaly ($\cA_n^agg$ diagrams where $g$ is the
graviton) and the cubic anomaly ($\cA_n^a\cA_n^b\cA_n^c$ diagrams)
are generically unrelated. There can also be more than one anomalous
$U(1)$ symmetry in the observable sector.

A radical new feature would have been the possibility of having
mixed $U(1)$ gauge anomaly coefficients ($\cA_n^aA_nA_n$ diagrams)
depending on the gauge group factor within $\cH_n$. However, as we
have shown these anomaly coefficients are always gauge--factor
independent in heterotic string--theory or M--theory.
In our more general five--dimensional approach, adopted in section 4,
we have, however, found that such a possibility can be realized.
Although the associated five--dimensional models are anomaly--free by
construction, it is not yet clear whether or not they are part of
M--theory. It is conceivable that they can be obtained
in the context of heterotic M--theory vacua with
five--branes~\cite{nse,dlow,dlow1,lows}.

\subsection{Special classes of compactifications}

So far, we have discussed the generic structure of anomaly
cancellation. This is the structure that arises without any further
information about the details of the compactification. Within a 
given class of compactifications, however, the information can be
much more precise. Technically speaking, for such a class of
compactifications one typically has some additional information about
the topological numbers that determine the anomaly coefficients via
the relations~\eqref{cr1}--\eqref{cr2}. This in turn, restricts the
structure of the anomaly coefficients and, hence, the structure of
anomalous $U(1)$ symmetries within the given class.

As an illustration of this, let us present an example.
A special role within heterotic M--theory is played by vacua for which
the topological numbers $\b_i$, defined in Eq.~\eqref{bi}, vanish. Such
vacua are also called symmetric. Recently, it has been shown~\cite{symm}
that such vacua based on Calabi--Yau three--folds indeed
exist~\footnote{It should be said that the examples of
Ref.~\cite{symm} do not directly apply to the present situation since
they were restricted to Calabi--Yau spaces with $h^{1,1}=1$. However,
one expects symmetric vacua to exist for some Calabi--Yau spaces with
$h^{1,1}>1$.}. The characteristic property of low--energy theories
based on symmetric vacua is the vanishing of the one--loop corrections
to the four--dimensional effective action. In our context,
we can use Eq.~\eqref{cr1}--\eqref{cr6} to derive the special anomaly
structure associated to symmetric vacua. We simply have to set
$\b_i=0$ in those equations. As the most remarkable simplification, we
then find from Eq.~\eqref{cr2} that the mixed gauge anomalies
($\cA_n^aA_nA_n$ diagrams) vanish in this case. That is, within effective
theories originating from symmetric vacua, the $U_I(1)$ symmetries of
type I may have at most mixed gravitational and cubic anomalies.

\subsection{Fayet--Illiopolous terms}

An important question in the context of anomalous $U(1)$ symmetries concerns the
possible associated FI terms. The structure that we have found is sufficiently
different from the usual one to reanalyze the question.

As usual, the FI terms can be read off from the $D=4$, $\cN =1$ 
K\"ahler potential. The moduli superfields are defined in terms of
their bosonic components as $S=V+i\sqrt{2}\s$ and
$T^i=a_{10}^i+i\sqrt{2}\c^i$. With these definitions, using the
kinetic terms~\eqref{S4kind} along with Eq.~\eqref{Sigma} and
\eqref{Xi1}, we find for the moduli K\"ahler potential $K$
\bea
 \k_4^2\, K&=& -\ln\left[\frac{1}{6}d_{ijk}\left( T^i+\bar{T}^i-
              \frac{\a '}{v^{1/3}}\eta_A^i\cA^A\right)
              \left(T^j+\bar{T}^j-\frac{\a '}{v^{1/3}}
              \eta_B^j\cA^B\right)\left(T^k+\bar{T}^k-
              \frac{\a '}{v^{1/3}}\eta_C^k\cA^C\right)\right]\nn \\
           && -\ln\left(S+\bar{S}+\frac{\k_4^2c^3}{2\p^2\a '}
              \cC_A\cA^A\right)\; .
\eea
Recall that the indices $A=(na)$ run over the two sectors as well as
over the various $U_I(1)$ symmetries in each sector.
Let us first focus on the $T$ part of this K\"ahler
potential. Expanding to linear order in the gauge fields, we find that
the FI terms from this part are proportional to
\begin{equation}
 d_{ijk}\eta_A^ia^ja^k
\end{equation}
where $a^i\sim a_{10}^i$ are the $(1,1)$ moduli. We have seen in
Eq.~\eqref{cons2}, however, that the $(1,1)$ moduli are constrained in
such a way that these expressions vanish~\cite{dsww}. Hence, we
conclude that the $T$ part of the K\"ahler potential does not lead to
FI terms.

What about the dilaton part? We find
\begin{equation}
 S_{\rm FI} =-\sum_{n=1}^2\int_{M_4}\sqrt{-g}\,\x_{na}D_n^a
\end{equation}
where the coefficients $\x_A$ are given by
\begin{equation}
 \x_{na} = \frac{c^3M^2g_0^2({\rm Re}(S))^{-1}}{16\p^2}C_{na}\; .\label{x}
\end{equation}
Here $M$ is the reduced Planck mass with $M^2=\k_4^{-2}$. The anomaly
coefficient $C_{na}$ is related to the mixed $U(1)$ gauge anomaly
($\cA_n^aA_nA_n$ diagrams). This coefficient has been generally defined in
Eq.~(\ref{cd2}b) and its specific form has been given in Eq.~\eqref{cr2}.
The result~\eqref{x} is very reminiscent of the one usually given
in the literature~\cite{dsw,ads,dis,dl}. However, there are some crucial
differences. Usually, it is assumed that all anomaly coefficients have
to be in fixed proportions to one another. The FI term can then be
expressed in terms of either one of them. Here we have seen that the
anomaly coefficients for the mixed gauge, the mixed gravitational and
the cubic anomaly are, in fact, generally different. What we have
found is that it is the mixed gauge anomaly coefficient that
determines the size of the FI term. We have also seen that, after a
suitable choice of basis, there can be one $U(1)$ symmetry with
a mixed gauge anomaly per sector. This means that we potentially have
two FI terms, one for the observable the other one for the hidden
sector. In particular, one could have a situation with only a
hidden sector FI term. The flexibility gained this way could
conceivably be useful for some purposes such as D--term inflationary
model building.

\subsection{Threshold corrections}

Another related question concerns the threshold corrections to the
gauge kinetic function, particularly for the $U_I(1)$ fields.
These can be read off from the action~\eqref{S4top}. The relevant part
of the action can be written in the form
\begin{equation}
 S_{\rm threshold} = -\frac{\e_S}{4g_0^2}\int_{M_4}{\rm Im}(T^i)\left(
                     -\b_i\tr F_1^2+\b_i\tr
                     F_2^2+f_{iAB}\cF^A\cF^B\right).
\end{equation}
where the strong coupling expansion parameter $\e_S$ is given by
\begin{equation}
 \e_S = \left(\frac{\k}{4\p}\right)^{2/3}\frac{2c\p^2\r}{v^{2/3}}\; .
\end{equation}
This implies the usual result for the gauge kinetic functions of the
$\cH_n$ parts of the gauge group namely
\begin{equation}
 f_n = S\mp\e_S\b_iT^i\; .
\end{equation}
For the $U_I(1)$ fields of type I we find
\begin{equation}
 f_{AB} = S\,\d_{AB}+\e_Sf_{iAB}T^i\; .
\end{equation}
The coefficients $f_{iAB}$ were explicitly defined in
Eq.~\eqref{fdef}. Note that the index $A=(na)$ runs over the two
sectors as well as the various $U_I(1)$ symmetries. Generally, the
coefficients $f_{iAB}$ have off--diagonal pieces that mix type I
$U_I(1)$ fields from the observable and the hidden sector.

\vspace{0.4cm}

{\bf Acknowledgments} 
A.~L. would like to thank Graham Ross for discussions. A.~L.~is
supported by the European Community under contract No.~FMRXCT 960090.
The work of K.S.S. was supported in part by PPARC under SPG grant 613.

\vspace{1cm}

%%%%%%%%%%%%%%%%%%%%%%%%%%%%%%%%%%%%%%%%%%%%%%%%%%%%%%%%%%%%%%%%%%%%%%%%%%%%

\appendix{\Large \bf Appendix}
\renewcommand{\theequation}{\Alph{section}.\arabic{equation}}
\setcounter{equation}{0}

%%%%%%%%%%%%%%%%%%%%%%%%%%%%%%%%%%%%%%%%%%%%%%%%%%%%%%%%%%%%%%%%%%%%%%%%%%%%

\section{Calculation of anomaly coefficients}

%%%%%%%%%%%%%%%%%%%%%%%%%%%%%%%%%%%%%%%%%%%%%%%%%%%%%%%%%%%%%%%%%%%%%%

In this appendix, we would like to be more explicit about the
calculation of the triangle anomaly
coefficients (\ref{c1}a--\ref{c6}f) using standard
technology~\cite{w1,gsw2}. We start by listing some useful trace
properties. For an $E_8\times E_8$ gauge field $F$ we have
\begin{equation}
 \Tr (F^6) = \frac{1}{48}\Tr (F^2)\Tr (F^4)-\frac{1}{14400}\left(\Tr
           (F^2)\right)^3
\end{equation}
where $\Tr$ is the trace in the adjoint and we define as usual
$\tr = \Tr /30$. Using a standard trick, we write
$F=\a T+\b\bar{F}$ where $T$ and $\bar{F}$ are in $E_8\times E_8$ and
$\a$ and $\b$ are arbitrary coefficients. Inserting this into the
above formula and extracting the $\a^3\b^3$ term we find
\bea
 \Tr (T^3\bar{F}^3) &=& \frac{1}{240}\Tr (T\bar{F}^3)\Tr
                   (T^2)+\frac{1}{240}\Tr (\bar{F}^2)\Tr
                   (\bar{F}T^3)+\frac{1}{80}\Tr (T\bar{F})
                   \Tr (T^2\bar{F}^2)\nn \\
                   &&-\frac{1}{24000}\Tr (T^2)\Tr (T\bar{F})\Tr
                   (\bar{F}^2)-\frac{1}{36000}\left(\Tr (T\bar{F})\right)^3\; .
                   \label{tr1}
\eea
Furthermore we need the $E_8$ relation
\begin{equation}
 \Tr (F^4) = \frac{1}{100}\left(\Tr F^2\right)^2
\end{equation}
where $F$ is now an $E_8$ gauge field. Inserting $F=\a T+\b\bar{F}$, as
above, where $T$ and $\bar{F}$ are now in $E_8$ we find from the
$\a\b^3$ and the $\a^2\b^2$ terms
\bea
 \Tr (T\bar{F}^3) &=& \frac{1}{100}\Tr (T\bar{F})\Tr (\bar{F}^2) \label{tr2}\\
 \Tr (T^2\bar{F}^2) &=& \frac{1}{150}\left(\Tr (T\bar{F})\right)^2
                        +\frac{1}{300}\Tr (T^2)\Tr (\bar{F}^2)\; .
 \label{tr3}
\eea
The above relations are useful to express the anomaly coefficients in
terms of compactification data. To see this let us define the generic
coefficient
\begin{equation}
 \cC (T) = \sum_r N_n^r\,\tr_{L_n^r} (T^3)
\end{equation}
for some $E_8\times E_8$ generator $T$. Using the index
theorem~\eqref{index} this can be put into the form
\begin{equation}
 \cC (T) =\frac{1}{6(2\p )^3}\int_X\left[\Tr
          (T^3\bar{F}^3)-\frac{1}{8}\Tr(T^3\bar{F})\tr\bar{R}^2\right]\; ,
\end{equation}
where $\bar{F}$ and $\bar{R}$ are the internal Yang--Mills and gravity
curvatures. We can rewrite this expression using
Eq.~\eqref{tr1}. Then choosing the generator $T$ to be part of one of
the $E_8$ factors, that is $T=T_n$, and using Eqs.~\eqref{tr2} and
\eqref{tr3} we can simplify this further. Finally, with the definition
of the topological numbers and the internal gauge fields given in
section 2.2 we find for the anomaly coefficient
\begin{equation}
 \cC (T_n) = \frac{3}{8\p}\left[\mp\tr (T_n^2)\tr
             (T_nQ_n^a)\eta_{na}^i\b_i+\frac{1}{12\p^2}\tr (T_nQ_n^a)
             \tr (T_nQ_n^b)\tr (T_nQ_n^c)\eta_{na}^i
             \eta_{nb}^j\eta_{nc}^k d_{ijk}\right]\; .
\end{equation}
A similar calculation can be carried out for the coefficient
relevant for the mixed gravitational anomaly defined by
\begin{equation}
 \cC^{(L)}(T_n) = \sum_rN_n^r\,\tr_{L_n^r}(T_n)\; .
\end{equation}
Then, choosing $T_n$ appropriately in these formulae, one derives
the various anomaly coefficients (\ref{c1}a--\ref{c6}f).

%%%%%%%%%%%%%%%%%%%%%%%%%%%%%%%%%%%%%%%%%%%%%%%%%%%%%%%%%%%%%%%%%%%%%%%%%%%%%%

%%%%%%%%%%%%%%%%%%%%%%%%%%%%%%%%%%%%%%%%%%%%%%%%%%%%%%%%%%%%%%%%%%%%%%%%%%%%

\end{document}